\pdfoutput=1

\documentclass[authoryear,preprint,12pt]{elsarticle}

\usepackage{amssymb}

\usepackage{bm}
\usepackage{lscape}
\usepackage{setspace}

\journal{Computers and Fluids}

\begin{document}

\begin{frontmatter}


\title{Combined effects of amplitude, frequency and bandwidth on wavepackets in laminar turbulent transition}
\author[ngs]{Kean Lee Kang\corref{cor1}}
\ead{kang_kl@u.nus.edu}
\cortext[cor1]{Corresponding author}
\author[mpe]{K. S. Yeo\corref{cor2}}
\cortext[cor2]{This article appeared in K.L. Kang and K.S. Yeo, Comput. Fluids (2019), https://doi.org/10.1016/j.compfluid.2019.104358. \\ \textcopyright 2019. This manuscript version is made available under the CC-BY-NC-ND 4.0 license http://creativecommons.org/licenses/by-nc-nd/4.0/.}
\address[ngs]{NUS Graduate School for Integrative Sciences and Engineering, National University of Singapore,
Singapore 117456}
\address[mpe]{Department of Mechanical Engineering, National University of Singapore, Singapore 117576}


\begin{abstract}
This study concerns wavepackets in laminar turbulent transition in a Blasius boundary layer. While initial amplitude and frequency have well-recognized roles in the transition process, the current study on the combined effects of amplitude, frequency, and bandwidth on the propagation of wavepackets is believed to be new. Because of the complexity of the system, these joint variations in multiple parameters could give rise to effects not seen through the variation of any single parameter. Direct numerical simulations (DNS) are utilized in a full factorial (fully crossed) design to investigate both individual and joint effects of variation in the simulation parameters, with a special focus on three distinct variants of wavepacket transition {\textemdash} the reverse Craik triad formation sequence, concurrent N-type and K-type transition and abrupt shifts in dominant frequency. From our factorial study, we can summarize the key transition trends of wavepackets as follows:
\begin{enumerate}
\item Broad bandwidth wavepackets predominantly transit to turbulence via the N-route. This tendency remains strong even as the frequency width is reduced.
\item Narrow bandwidth wavetrains exhibit predominantly K-type transition. The front broadband part of an emerging wavetrain may experience N-type transition, but this wavefront should not be considered as a part of truly narrow-bandwidth wavepackets. 
\item K-type transition is the most likely for wavepackets that are initiated with high energy/amplitude and/or with the peak frequency at the lower branch of the neutral stability curve.
\end{enumerate}
\end{abstract}


\end{frontmatter}


\section{\label{sec:intro}Introduction}
A Blasius boundary layer is known to transition to turbulence through several mechanisms, known as the K (Klebanoff), N (Novosibirsk) and oblique transition regimes, as classified by \citet{Boiko2012}. For individual waves of well-defined frequency, wavenumber and amplitude, much is already known about its transition route. For example, it is known that higher-amplitude waves are more likely to transition via the K-type (aligned $\Lambda$-structures/vortices) \citep{Boiko2012}, while those of lower amplitude are associated with N-type transition (staggered $\Lambda$-structures). The challenge now is perhaps to integrate these results into a more robust framework that is better able to predict transition under the vagaries and complexities of real-world conditions. More complex systems such as wavepackets still hold many unanswered questions. For example, the limited spatial extent of wavepackets means that they are unable to support truly 2D waves, and any aligned or staggered $\Lambda$-structure patterns may be distorted or truncated due to the limited spatial extent of the wavepacket. Moreover, the finite duration of the wavepacket also supports a broad frequency spectrum.

The development of wavepackets in a Blasius boundary layer was first investigated by \citet{Gaster1968, Gaster1975a}. A three-wave system known as the Craik triad \citep{Craik1971} was discovered to play a major role \citep{Smith1987, Cohen1991} in the nonlinear evolution of wavepackets and these triads are closely associated with N-type transition, which is also known as H-type transition \citep{Herbert1984,Herbert1988}. More recently, the mechanism of genesis and evolution of the spatio-temporal wavepacket in a linear framework was explored in \citet{Sengupta2006} and \citet{Sengupta2006a}, and the influence of frequency in determining either K-type or H-type transition in wavepackets was described in \citet{Bhaumik2014a,Bhaumik2015,Bhaumik2015a,Sharma2018}, who used the synonymous term ``spatio-temporal wave front'' (STWF) for a wavepacket. This spatio-temporal wavepacket is a modulated and band-limited combination of spatio-temporal eigenmodes of the Orr-Sommerfeld equation as described by Equation 1 in \citet{Bhaumik2014a}.

The Craik triad consists of a 2D mode known as the ``fundamental'' and two 3D oblique modes at half the frequency of the fundamental, known as the ``subharmonic''. The two 3D modes have spanwise wavenumbers of equal magnitude but opposite sign, such that they propagate at equal and opposite angles $\pm\theta$ from the streamwise direction. Later work by \citet{Herbert1984,Herbert1988} and \citet{Kachanov1984} showed that ``detuning'' can also occur, in which strongly nonlinear interactions occur even between modes that do not precisely fulfil Craik's triad conditions.

Subsequently, \citet{Medeiros1999a,Medeiros1999b} explored how adjustments to the frequency spectrum of the initial wavepacket affected its transition process. It was found that elimination of the subharmonic mode from the initial spectrum did not significantly alter the wavepacket's downstream development. This was because the subharmonic mode could be generated via nonlinear difference interactions within a fundamental band of sufficient width, according to a model proposed by \citet{Craik2001} which is supported by evidence from the experiments of \citet{Kachanov1984} and \citet{Paula2010}. This theory considers the fundamental not as a discrete mode, but as a band with a continuous distribution of modes within it. Different modes within the band can interact to produce the subharmonic mode.

The effect of wavepacket frequency bandwidth was investigated by \citet{Medeiros2006}, who created wavepackets of different frequency bandwidth by introducing a controlled temporal modulation of a wavetrain emanating from a point source in a boundary layer. The temporal modulation resulted in a controllable bandwidth, which in turn produced a streamwise modulation when a wavepacket evolved downstream. The term ``wavetrain'' is used for narrow bandwidth disturbances because of the inverse relationship between size in spectral space and physical space. Intuitively, it can be said that the more concentrated a function is in Fourier space, the more spread out it will be in physical space.\footnote{This mathematical principle is famously applied in the Heisenberg uncertainty principle, in which position and momentum are a Fourier transform pair to within a factor of Planck's constant.}

In accordance with \citet{Craik2001}, it was found that broad bandwidth wavepackets developed a subharmonic mode, but narrow bandwidth wavetrains did not. Instead, the narrow bandwidth disturbances formed Klebanoff modes (low frequency modes which are physically longitudinal vortical and streaky structures) \citep{Wu2001,Goldstein2008}. Interestingly, a mean flow distortion was also observed in the wavepackets, even though it was not introduced at the source, and it was suggested to be the ``source'' of the subharmonic band's energy \citep{Medeiros2006}. 

\citet{Bhaumik2017a} and \citet{Sharma2018} found that the modulation characteristics and consequently the features of the wavepacket change when one shifts from an impulsive monochromatic onset of excitation to a gradual monochromatic excitation. In the case of switching the exciter on and off (finite onset and termination of the exciter), it would result in the evolution of two subsequently phase-shifted and time-delayed spatio-temporal wavepackets as noted in \citet{Bhaumik2013} and \citet{Sundaram2019} for 2D disturbances - one wavepacket when the exciter is switched on, and another when the exciter is switched off, while the spatial Tollmien-Schlichting (TS) wavepacket/wavetrain recedes back.   

Furthermore, \citet{Paula2013} performed experiments on an airfoil boundary layer showing that the steepness of the modulation of a 2D wave does not play any significant role in the transition process. Instead, basic nonlinear interactions between modes of a modulated 2D wave generate quasi-subharmonic modes that significantly affect the transition scenario. \citet{Paula2013b} showed experimentally that low-frequency waves are produced by difference interactions among waves in a narrow band of frequencies about the fundamental modes, supporting the hypothesis of \citet{Craik2001}.

\section{Methods}
\subsection{\label{subsec:dns-code} DNS code}
The full details of the direct numerical simulation (DNS) code used in this work have been published in \citep{Wang2003, Wang2005}. Here, we give an outline of the code for application to our specific needs.

The Navier-Stokes equation is implemented in the code as the momentum equation in the nonlinear perturbation form,
\begin{equation}
\frac{\partial\mathbf{u}}{\partial t}+(\mathbf{U}\centerdot\nabla)\mathbf{u} +(\mathbf{u}\centerdot\nabla)\mathbf{U} +(\mathbf{u}\centerdot\nabla)\mathbf{u} =-\nabla p + \frac{1}{Re}\nabla^2 \mathbf{u}.
\label{eq:dns-momentum-equation}
\end{equation}
where $\mathbf{u}$ and $\mathbf{U}$ are the perturbation and steady-state (time independent) velocity vector fields respectively, $p$ is the perturbation pressure scalar field and $Re$ is the Reynolds number. In this paper, $\mathbf{U}$ is a Blasius mean flow solution.

Second-order finite volume spatial discretization and second-order backward Euler temporal discretization of equation (\ref{eq:dns-momentum-equation}) is then applied. Our time splitting strategy is a fractional step. This highly efficient technique was developed by \citet{Chorin1969}, \citet{Temam1984} and \citet{Kim1985}, and it has emerged as one of the most popular DNS algorithms in use today. For numerical stability, we use a fully implicit iterative variant of the fractional step method with a pressure correction scheme \citep{Wesseling2001} \citep[p. 118]{Holmes2012}. At each time step, outer iterations yield an intermediate velocity vector field that is updated by a corrective pressure scalar field from inner iterations. Spatially, the DNS code adopts finite volume discretization on a collocated grid system formulated in general curvilinear nonorthogonal coordinates. Collocation of the velocity and pressure data at the same grid points triggers numerical perturbation pressure oscillations that are stabilized by the momentum interpolation method of \citet{Rhie1983}. A geometric multigrid procedure \citep{Wesseling2001} is employed to solve the pressure-Poisson problem, with an alternating direction implicit (ADI) solver \citep{Birkhoff1962} in 3D as the smoother and the full approximation storage (FAS) algorithm of \citet{Brandt1977} used. 

\subsection{\label{subsec:dns-domain} Computational domain and parameters}
The following computational domain and parameters are modelled after the experimental setup of \citet{Cohen1991}. While the DNS code itself is formulated in general curvilinear coordinates, this article will use Cartesian coordinates, which are sufficient to investigate our simple domain geometry of a flat plate boundary layer. The streamwise, wall-normal and spanwise Cartesian coordinates are denoted by $x$, $y$ and $z$ respectively. They are non-dimensionalized based on the reference length $\delta_0=2.3182\times 10^{-3}$ m, which is the boundary layer displacement thickness at the disturbance source. If $x^*$, $y^*$ and $z^*$ are dimensional lengths, then $x=x^*/\delta_0$, $y=y^*/\delta_0$ and $z=z^*/\delta_0$. The computational domain is a box with $310\leq x \leq 1510$, $0\leq y \leq 54$, $-173\leq z \leq 173$, and it is meshed with $1186\times 85\times 195$ grid points, with uniform meshing in the $x$ and $z$ direction, and a stretched grid in the $y$ direction to increase grid resolution close to the wall, according to the formula
\begin{equation}
y=\frac{y_{max}\gamma\xi}{\gamma\xi_{max}+y_{max}\left(\xi_{max}-\xi\right)},
\end{equation}
where $\xi$ is the index number of the grid point. Thus, $\xi$ is an integer satisfying $0< \xi \leq \xi_{max}=85$. Similarly, $y$ will be a real number such that $0\leq y \leq y_{max}=54$. $\gamma$ is a stretching parameter that is set to 1.6. Out of these 85 grid points in the wall-normal direction, 54 are located within the boundary layer at the inflow of the domain, according to the $0.99U_\infty$ criterion. (Given that the boundary layer thickness increases on average as we progress in the streamwise direction, 54 also represents an approximate lower bound on the number of grid points across the boundary layer in the rest of the domain.)

Freestream velocity is $U_\infty = 6.65$ m/s and kinematic viscosity is $\nu = 1.49\times 10^{-5}$ ${\textrm{m}}^2/{\textrm{s}}$. The disturbance source is located at $x=349.4$, giving rise to a displacement thickness Reynolds number ${Re}=\delta U_\infty /\nu = 1035$, or in terms of momentum thickness Reynolds number ${Re}_\theta=\theta U_\infty /\nu = 399$. The Reynolds numbers at the inflow and outflow of the DNS domain are ${Re}=975$ (${Re}_\theta=376$) and ${Re}=2151$ (${Re}_\theta=830$) respectively. Time is non-dimensionalized as $t=t^*U_\infty /\delta_0$, with $t^*$ measured in seconds. The non-dimensional angular frequency is $\omega = 2\pi f\delta_0 /U_\infty$, where $f$ is the frequency in Hertz. The symbols $u$, $v$ and $w$ represent the streamwise, wall-normal and spanwise perturbation velocities respectively, on a Blasius mean flow profile. These velocities are also non-dimensional; for example, $u\equiv u^*/U_\infty$ where $u^*$ is the dimensional velocity.

The inflow boundary condition is zero perturbation velocity, which is equivalent to a laminar boundary layer inflow. At the wall, the no slip condition is imposed, while far from the wall, the perturbation velocity is assumed to be zero, corresponding to freestream conditions. Periodic boundary conditions are used in the spanwise direction. At the outflow boundary, the streamwise second derivative of all velocity components is set to zero; $\partial^2u/\partial x^2=\partial^2v/\partial x^2=\partial^2w/\partial x^2=0$. A buffer domain region \citep{Liu1994} is also implemented just before the outflow boundary (between $1507.6\leq x \leq 1510$, corresponding to dimensional range $3.495\textrm{ m}\leq x^* \leq 3.5\textrm{ m}$), to prevent wave reflections upstream. There are two buffer functions being used on the diffusion terms of the governing equations within the buffer domain. The first buffer function acts only on the streamwise diffusion terms to slowly parabolize the governing equations in the streamwise direction, while the second buffer function applies to all the diffusion terms in order to reduce the Reynolds number in the flow to below the critical Reynolds number such that all disturbance modes are damped. Additional details regarding the implementation of the buffer domain, including the mathematical buffer domain function specifications, are given in \citet{Wang2005}. 

\subsection{\label{subsec:source-details} Wavepacket source details}
Our disturbance source is a wall-normal perturbation velocity specified within a circle on the wall. To minimize numerical artefacts, the disturbance is imposed in a spatially smooth manner, distributed over a few grid points with a Gaussian envelope along the wall. In particular, the initial disturbance is applied to grid points $(x,z)$ on the wall satisfying $\sqrt{(x-x_0)^2+(z-z_0)^2}<R$ where $R=\sqrt{8}$ is the radius of a circular disk centered at $(x_0,z_0)=(349.4,0)$. The input disturbance function $v(t)$ is multiplied by a two-dimensional Gaussian function such that within the disk:

\begin{equation}
v_{source}(x,z,t)=v(t) \exp\left[-\frac{(x-x_0)^2}{2}-\frac{(z-z_0)^2}{2}\right].
\end{equation}

The input disturbance is $v(t) = A\sin(\omega_0t)$, where $A$ is the amplitude and $\omega_0$ is the source driving frequency. The values used for the simulations are given in Table~\ref{tab:factorial-sims}. 

\subsection{\label{subsec:sim-design} Simulation design}
A full factorial-type table of direct numerical simulation (DNS) runs was used, with the full details of the initial disturbances given in Table~\ref{tab:factorial-sims}. The cases are named according to the following convention. Each case is designated by a name of the form \_E-\_F-\_B where the letter E represents energy or power, F is angular frequency $\omega_{max}$ (the frequency with the highest spectral density) and B is 3dB-bandwidth $\Delta\omega$.\footnote{Decibels express a ratio comparing the power of one value with another on a base-10 logarithmic scale. Therefore, the 3dB bandwidth refers to the frequency width (expressed in terms of $\Delta f$ or $\Delta\omega$) in which the spectral density is greater than or equal to half its maximum value ($10^{0.5}\approx3$).} The letters L, M and H denote low, medium and high respectively, and are filled into the blanks. So LE-MF-HB is a case with low energy, medium frequency and high (broad) bandwidth. 

The base parameters were chosen to match the experiments of \citet{Cohen1991} for the LE-MF-HB case. Subsequently, amplitude was adjusted to maintain approximately the same total energy of the initial disturbance in the other LE cases, under the simplifying assumption of a sinusoidal wave for the initial disturbance with power proportional to the square of the amplitude $P \propto A^2$. The HE cases used double the amplitude of the LE cases. Bandwidth is controlled indirectly by varying the duration of the initial disturbance (the length of time the source is switched on, beginning at $t=0$). Broad bandwidth wavepackets are excited for one period $\frac{2\pi}{\omega_0}$, medium bandwidth wavepackets are excited for four periods  $4(\frac{2\pi}{\omega_0})$ and wavetrains are excited continually (for the entire duration of the simulation). For example, if $A$ represents amplitude and $T_d$ the disturbance source excitation duration, while subscript ``a'' denotes the broadband single cycle pulse, and subscript ``b'' denotes the medium bandwidth four cycle pulse, we can match the total energy, $E$ of the input signal by setting $E_b=E_a$.
\begin{equation}
T_{db}A^2_b=T_{da}A^2_a
\Longrightarrow
4\left(\frac{2\pi}{\omega_b}\right)A^2_b=
\left(\frac{2\pi}{\omega_a}\right)A^2_a
\Longrightarrow
A_b=\frac{A_a}{2}\sqrt{\frac{\omega_b}{\omega_a}},
\label{eq:amplitude-calculation}
\end{equation}
where $\omega_a$ and $\omega_b$ are the driving frequencies of the single and four cycle pulses respectively.

\begin{figure}
\centering
\includegraphics{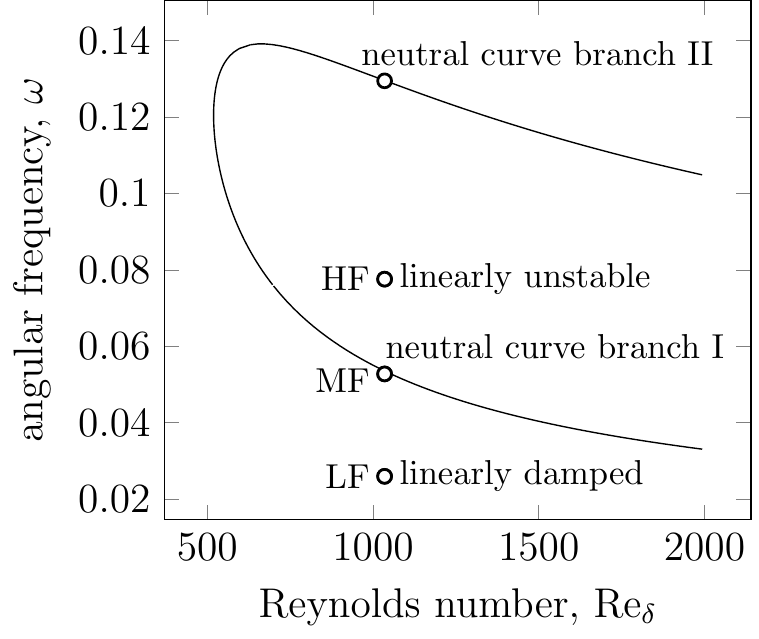}
\caption{\label{fig:neutral-curve}Locations of initial wavepacket frequencies relative to the linear stability theory neutral curve. All these points are at the displacement thickness Reynolds number, $\mathrm{Re}_\delta$ of 1035, which corresponds to the disturbance source location.}
\end{figure}

We obtain the neutral curve using a linear stability code following the methodology of \citet{Jordinson1970}. Then we select the angular frequency with maximum spectral density $\omega_{max}$ based on the locations of the various disturbances relative to the neutral curve as may be seen in Figure~\ref{fig:neutral-curve}.\footnote{The driving frequency of the source is not the same as the frequency of the peak spectral density $\omega_{max}$ in the wavepacket due to the finite support of the signal in time. The source driving frequency has been carefully tuned to achieve the specified $\omega_{max}$ in the resulting wavepacket. More details can be found in \citet{Kang2013}.} The medium frequency, MF case corresponds to branch I of the neutral stability curve at the disturbance source. The high frequency HF case is the mode with a high spatial growth rate, $\alpha_i$ (most linearly unstable) and the low frequency LF is in the linearly damped region.

Testing three different bandwidths at three different frequencies requires $3^2=9$ simulation runs. Thereafter, these nine cases are run again at double their respective amplitudes (four times the energy or power), yielding a total of $9\times2=18$ simulations. 

{\setstretch{1.0}
\begin{table}
  \begin{center}
  \begingroup
  \fontsize{10pt}{12pt}\selectfont
  \begin{tabular}{p{0.12\textwidth}p{0.22\textwidth}p{0.14\textwidth}p{0.17\textwidth}p{0.17\textwidth}}
  \hline
Amplitude & Angular frequency with max spectral density, $\omega_{max}$ & Source driving frequency, $\omega_0$ & 3dB-bandwidth, $\Delta\omega$ & Case name\\
\hline
       0.0141   & 0.0259 & 0.0309 & 0.0292 & LE-LF-HB\\
       0.0054   &            & 0.0262 & 0.0058 & LE-LF-MB\\
       0.0168   &            & 0.0259 & 0.0019 & LE-LF-LB\\ \hline
       0.0082   & 0.0528 & 0.0630 & 0.0594 & LE-MF-HB\\
       0.0038   &            & 0.0533 & 0.0117 & LE-MF-MB\\  
       0.0082   &            & 0.0528 & 0.0039 & LE-MF-LB\\ \hline
       0.0047   & 0.0776 & 0.0927 & 0.0874 & LE-HF-HB\\
       0.0028   &            & 0.0783 & 0.0172 & LE-HF-MB\\
       0.0056   &            & 0.0777 & 0.0057 & LE-HF-LB\\ \hline\hline
       0.0282   & 0.0259 & 0.0309 & 0.0292 & HE-LF-HB\\
       0.0108   &            & 0.0262 & 0.0058 & HE-LF-MB\\
       0.0336   &            & 0.0259 & 0.0019 & HE-LF-LB\\ \hline
       0.0138   & 0.0528 & 0.0630 & 0.0594 & HE-MF-HB\\
       0.0082   &            & 0.0533 & 0.0117 & HE-MF-MB\\  
       0.0164   &            & 0.0528 & 0.0039 & HE-MF-LB\\ \hline
       0.0094   & 0.0776 & 0.0927 & 0.0874 & HE-HF-HB\\
       0.0056   &            & 0.0783 & 0.0172 & HE-HF-MB\\
       0.0112   &            & 0.0777 & 0.0057 & HE-HF-LB\\ \hline
        \hline
  \end{tabular}
  \endgroup
  \caption{\label{tab:factorial-sims}Parameters of the initial disturbance used to generate the wavepacket in each case.}
  \end{center}
\end{table}
}

The DNS simulations investigated 18 combinations of wavepacket frequency, amplitude and bandwidth to elucidate the complex interplay among these factors that determines the transition path of the wavepacket. Most of the wavepackets break down to incipient turbulence within the domain, and this enabled us to have a fuller picture of the laminar-turbulent transition process. 

\subsection{\label{subsec:dns-convergence} DNS grid convergence}
To check the adequacy of the grid resolution, the broadband wavepacket case was also run at a higher resolution. This was done by first determining the spacing between points of the original $1186\times 85\times 195$ grid: $\Delta x^+=24.554$, $\Delta y^+_{min}=0.4595$ and $\Delta z^+=43.128$. These spacings are given in terms of non-dimensional wall units using the standard formulae $\Delta x^+ = \Delta x^* u_\tau/\nu$ and $u_\tau = \sqrt{\tau_w/\rho}$ where $\Delta x^*$ is the dimensional grid spacing, $u_\tau$ is the friction velocity at the source, $\tau_w$ is the wall shear stress at the source and $\rho$ is the fluid density. Note that $\Delta y^+_{min}$ is the grid spacing in the wall-normal direction just above the wall. The largest grid spacing is in the spanwise $z$-direction, hence this was the target of the most aggressive grid refinement. Moreover, the z-resolution (spanwise direction) is crucial as \citet{Yeo2010} noted to accurately reflect the beta-expansion phase of transition. The refined grid has $1586\times 101\times 391$ points, or $\Delta x^+=18.39375$, $\Delta y^+_{min}=0.3851$ and $\Delta z^+=21.564$. In terms of temporal resolution, the original grid has non-dimensional time step $\Delta t=0.25$ giving rise to a Courant-Friedrichs-Lewy (CFL) number $u \Delta t/\Delta x =0.2472$, while the new grid has $\Delta t=0.2$ to yield CFL $u \Delta t/\Delta x =0.2645$. Comparing the time step with the initial characteristic period, the source driving frequency of $\omega_0=0.063$ means a period of $\frac{2\pi}{\omega_0}=100$. With a non-dimensional time step of $\Delta t\leq0.25$, we have $\geq400$ time steps within one period of the initial disturbance in both refined and standard resolution simulations. The results with the refined grid show no significant difference with the lower-resolution results.

\begin{figure}
\centering
\includegraphics{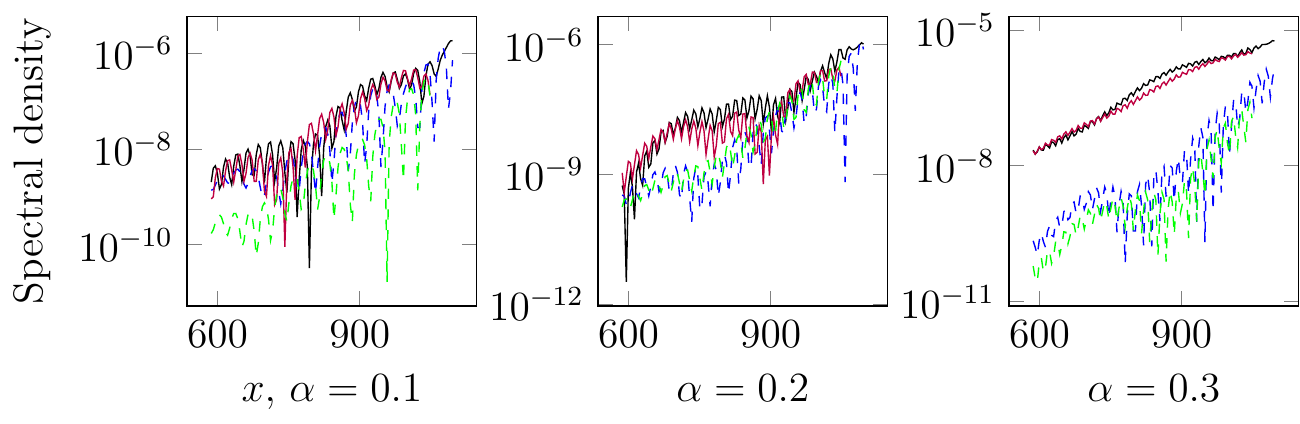}
\caption{\label{fig:single-cycle-modetracking}Spatial evolution of the broadband (LE-MF-HB) wavepacket fundamental mode $(\omega,\beta)=(0.056,0)$ (solid line) and subharmonic mode $(0.028,0.125)$ (dashed line) at different values of streamwise wavenumber $\alpha$. The results obtained using the standard-resolution DNS grid are shown in black and blue (darker colors) and are compared with data from a higher-resolution DNS shown in purple and green (lighter colors).}
\end{figure}

Figure~\ref{fig:single-cycle-modetracking} shows the convergence of spatial evolution results, with good agreement for the fundamental mode. While the subharmonic mode shows more variation, the overall trend is the same at both grid resolutions, bearing in mind that flow transition is a chaotic amplifying process, so that its details are sensitive to initial conditions.

This grid convergence test supplements previous grid convergence or validation studies done with the same DNS code over a very similar flow configuration \citep{Zhao2007, Yeo2010}. Numerical validation of the code had also been performed in \citet{Wang2005}. 

\subsection{\label{subsec:experiment-comparison}Comparison with experiments}
The present computational framework closely mimics the experimental setup of \citet{Cohen1991}. Good qualitative and quantitative agreement on the transitional and breakdown behavior of wavepackets had been obtained by \citet{Yeo2010} with \citet{Cohen1991}. Good qualitative agreement with the experimental results of \citet{Medeiros2006} and theory of \citet{Craik2001} was also observed in the present simulations as we shall see further on. For a start, Table~\ref{tab:cohen91fig8a-medeiros99bfig4-comparison} compares spectral data from the present study for the LE-MF-HB wavepacket (which is initiated by a small delta-like disturbance pulse) with the similar experimental results from \citet{Cohen1991,Medeiros1999b} – where generally good agreement may be noted with regards to the properties of the fundamental, the subharmonic and the first low frequency modes. 

\begin{table}
  \begin{center}
  \begin{tabular}{lp{0.2\textwidth}p{0.2\textwidth}p{0.2\textwidth}}
  \hline
  &Present work (LE-MF-HB) & Cohen et al. (1991) & Medeiros and Gaster (1999)\\ \hline
Sampling location  & \mbox{$2.60$ m} \mbox{($x=1122$)} & $2.60$ m & $1.30$ m\\ 
Sampling location $Re_\delta$ & 1853 & $\sim1800$ & $\sim2100$ \\
Fundamental mode  & $(0.100,0)$ & $(0.093,0)$ & (0.096,0)\\
Subharmonic mode & $(0.061,0.225)$ & $(0.053,0.271)$ & (0.065,0.185)\\
Low-frequency mode 1 & $(0,0.134)$ & $(0,0.131)$ & - \\
Low-frequency mode 2 & $(0,0.463)$ & $(0,0.619)$ & - \\ \hline
  \end{tabular}
  \caption{\label{tab:cohen91fig8a-medeiros99bfig4-comparison}Comparison of the $(\omega_\delta,\beta_\delta)$ location of the various modes in the 2D spectrum of the present work LE-MF-HB wavepacket with those in figure 8(a) of \citet{Cohen1991} and the $0^\circ$ case of figure 4 in \citet{Medeiros1999b}. In this table, the frequency-wavenumber pairs $(\omega_\delta,\beta_\delta)$ have been non-dimensionalized based on the \emph{local} displacement thickness $\delta$ in order to match the procedure used in \citet{Cohen1991} and \citet{Medeiros1999b}.}
  \end{center}
\end{table}

\subsection{\label{subsec:lst-comparison}Comparison with linear stability theory}
\begin{figure}
\includegraphics{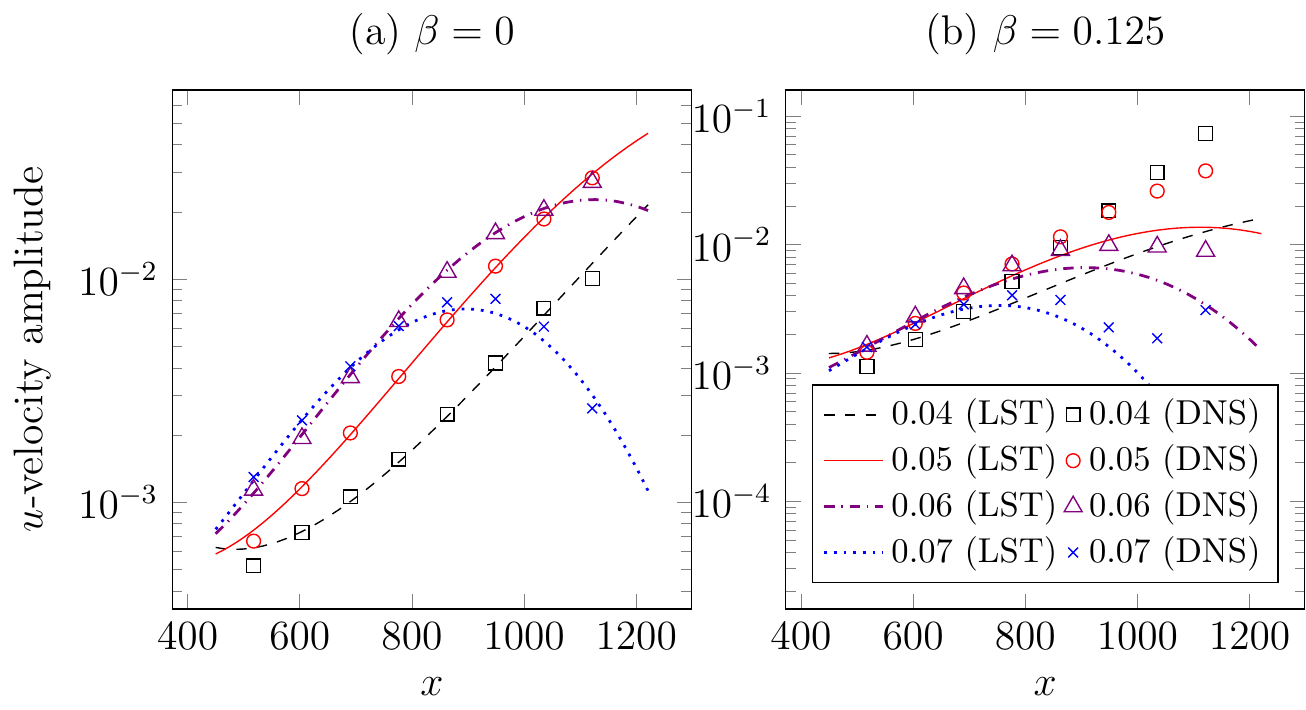}
\caption{\label{fig:lst-dns-comparison-broadband}Comparison between the streamwise development of waves in the DNS simulation (symbols) and linear stability theory (lines), for the broadband input pulse LE-MF-HB. The plot on the left is for spanwise
wavenumber $\beta=0$, while the right plot is for $\beta=0.125$. The angular frequencies $\omega$ are represented by: Line/symbol: dashed/square: $\omega=0.04$, solid/circle: $\omega=0.05$, dash-dotted/triangle: $\omega=0.06$, dotted/x: $\omega=0.07$.}
\end{figure}

To compare the development of various spectral components of the broadband LE-MF-HB wavepacket in DNS with the predictions of linear stability theory (LST), the amplitudes of several $(\omega,\beta)$ modes were extracted from the DNS simulation at various streamwise locations. The amplitude is calculated as the square root of the spectral density of the mode \citep[p. 1138]{Cohen1994}. This amplitude information was plot together in Figure~\ref{fig:lst-dns-comparison-broadband} with the corresponding spatial amplification curves predicted by linear stability theory (LST) \citep{Jordinson1970}. Specifically, the LST curves show the total amplification of a disturbance propagating downstream at constant frequency and spanwise wavenumber. The initial amplitude of each LST curve is chosen for best fit with the DNS data. The first data point is located slightly downstream of the source to give room for the nonlinear transients associated with the wavepacket generation process to die away.

From Figure~\ref{fig:lst-dns-comparison-broadband}, we note good agreement between the LST and wavepacket components during the early stage of wavepacket growth. The oblique waves tended to deviate fairly early due to the onset of nonlinear effects, but the 2D waves sustained agreement with the LST for a much longer distance, which had also been observed by \citet{Cohen1994}. Further comparisons with LST are given later in Figure~\ref{fig:test39sine} and \ref{fig:test39sine2}. We built on this foundation to perform our subsequent investigations by making changes to the amplitude, frequency and bandwidth of the initiating disturbance.

\section{Results}
Over the course of our investigations of the 18 wavepackets described in Table~\ref{tab:factorial-sims}, we came across several cases with special or unique behaviour, namely:
\begin{enumerate}
\item Reverse tried formation sequence
\item Concurrent N-type and K-type transition
\item Continuous and abrupt peak frequency shift
\end{enumerate}

The evidence for these will be presented and discussed first in Sections~\ref{sec:reverse-triad-sequence}, \ref{sec:mixed-transition} and \ref{sec:peak-freq-jump} respectively, followed by a summary of the overall trends that we have found across all 18 wavepackets in Section~\ref{sec:transition-types}.

\subsection{\label{sec:reverse-triad-sequence} Reverse triad formation sequence}
Here, we study the low-frequency wavepackets, cases LE-LF-LB (low energy, low frequency, narrow bandwidth) and LE-LF-MB (low energy, low frequency, medium bandwidth) which display a novel spectral behavior. These wavepackets are started by an initial disturbance with half the frequency of the medium frequency cases. In other words, they are at the frequency of the subharmonic in the MF wavepackets. The velocity contours are shown in Figure~\ref{fig:u-velocity-contours-test41test42test43-part1} and \ref{fig:u-velocity-contours-test41test42test43-part2}, with their associated spectral density plots in Figure~\ref{fig:u-spectra-contours-test41test42test43-part1} and \ref{fig:u-spectra-contours-test41test42test43-part2}. Note that for the narrowband case at $X=1294$ in Figure~\ref{fig:u-velocity-contours-test41test42test43-part2}, the periodicity breaks up by the end of the time domain, but this is likely to be a numerical artefact arising at the final breakdown to turbulence stage that does not affect this study's focus on the early and middle stages of transition.

\begin{landscape}

\begin{figure}
\centering
\includegraphics[width=\linewidth]{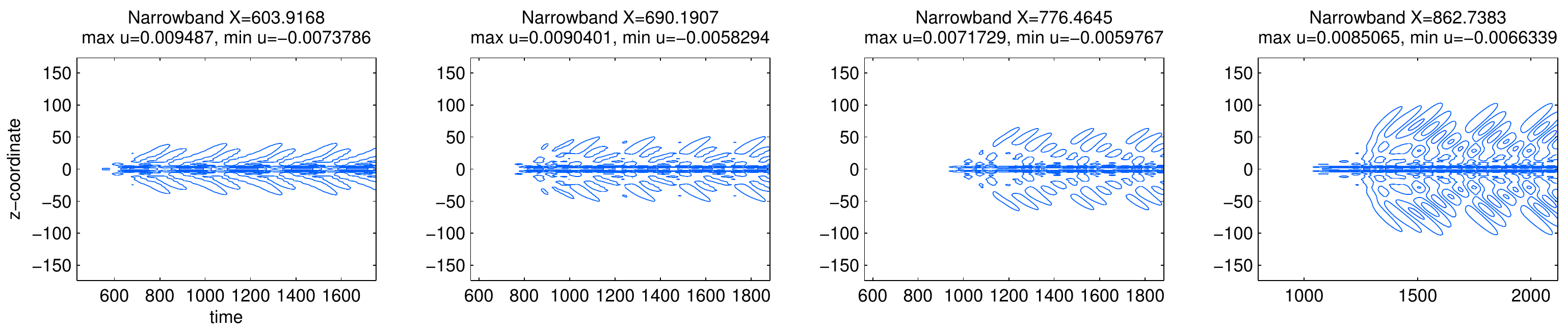}
\includegraphics[width=\linewidth]{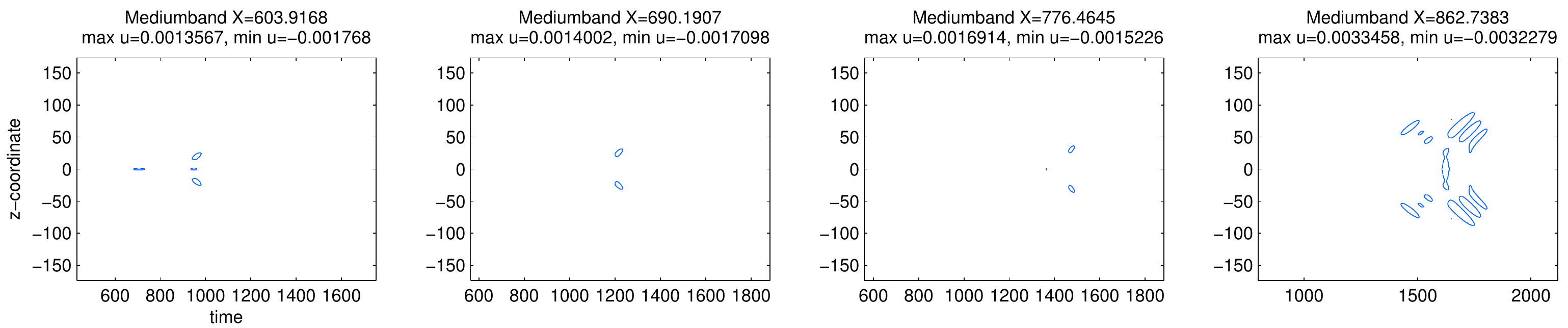}
\caption[]{\label{fig:u-velocity-contours-test41test42test43-part1}Contours of LE-LF (low energy low frequency) $u$-velocity data collected at $y^*/\delta=0.6$ and (left to right) $x=604$, $690$, $776$ and $863$. The first row is the narrow bandwidth, continually-excited wavetrain and the second row is the medium bandwidth input pulse. Both the contour lines and their colors have been standardized so comparison can be made between all the images in this figure and in Figure~\ref{fig:u-velocity-contours-test41test42test43-part2}. The contour lines are colored with a scale that varies linearly from dark blue ($u= -0.3$) to red ($u=0.8$).\\ \centerline{Colorbar: min \includegraphics[height=0.8\baselineskip]{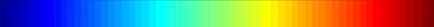} max}}
\end{figure}

\begin{figure}
\centering
\includegraphics[width=\linewidth]{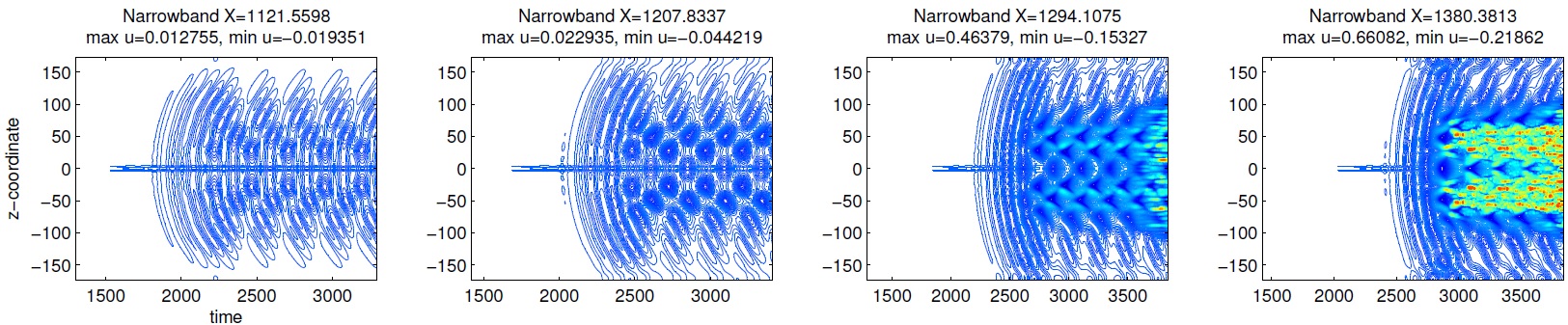}
\includegraphics[width=\linewidth]{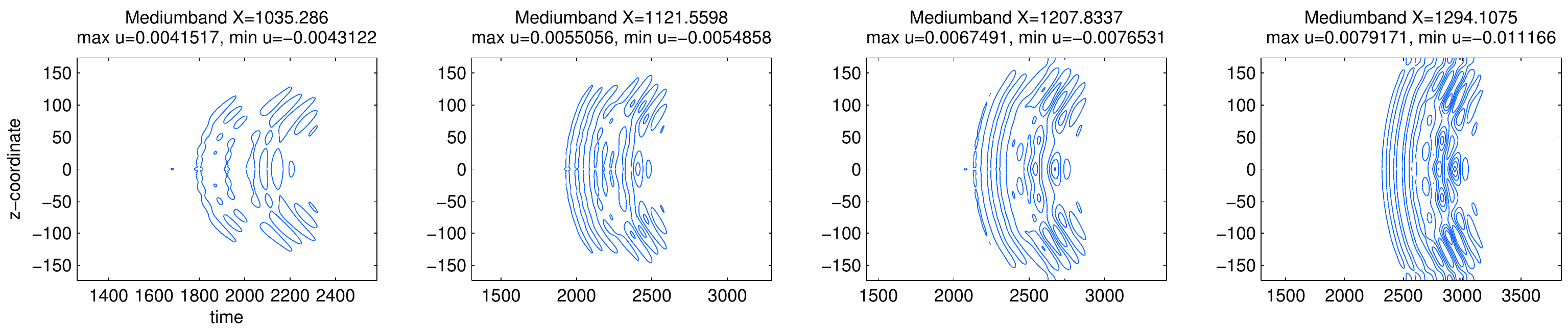}
\caption[]{\label{fig:u-velocity-contours-test41test42test43-part2}Contours of LE-LF (low energy low frequency) $u$-velocity data collected at $y^*/\delta=0.6$ and $x$ locations indicated above each plot. The first row is the narrow bandwidth, continually-excited wavetrain and the second row is the medium bandwidth input pulse. The contours and colors in these figures have been standardized so comparison can be made between all the images in this figure and in Figure~\ref{fig:u-velocity-contours-test41test42test43-part1}.\\ \centerline{Colorbar: min \includegraphics[height=0.8\baselineskip]{colormap-jet.jpg} max}}
\end{figure}

\begin{figure}
\centering
\includegraphics[width=\linewidth]{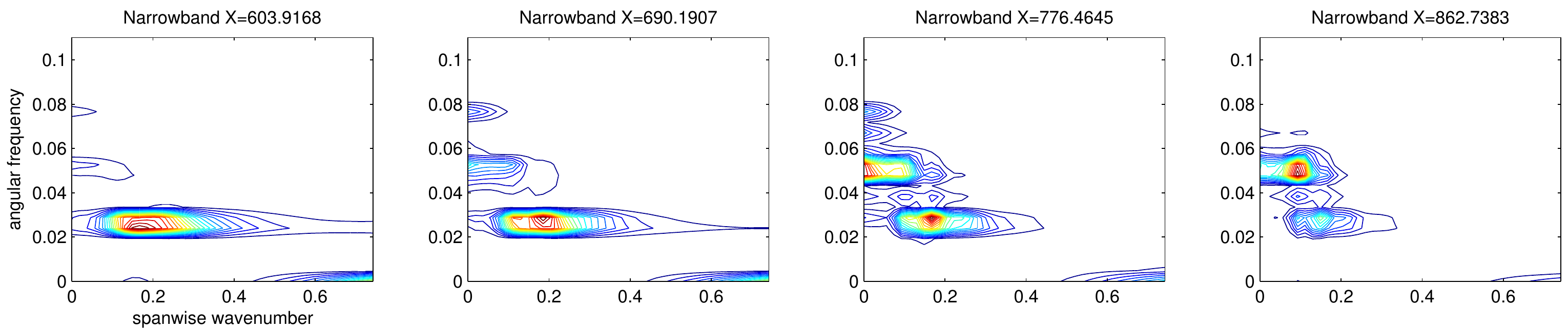}
\includegraphics[width=\linewidth]{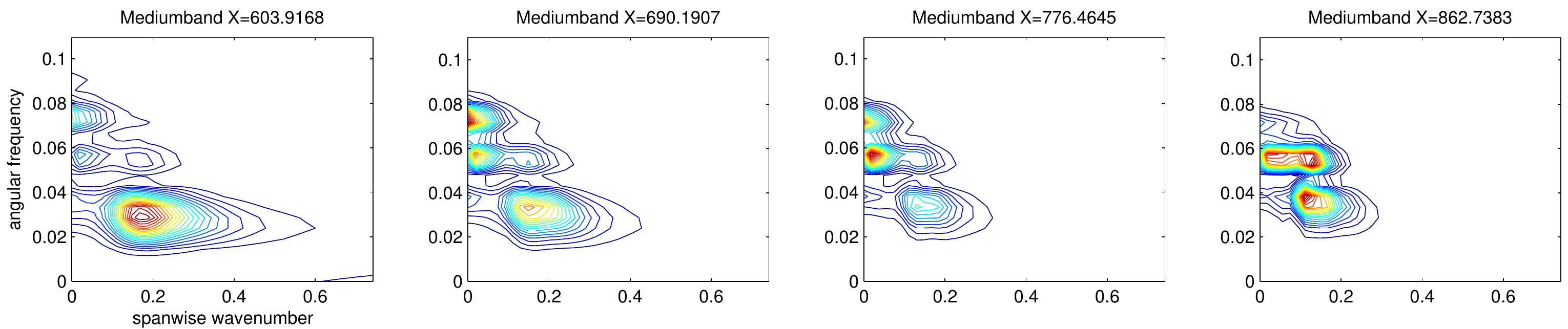}
\caption[]{\label{fig:u-spectra-contours-test41test42test43-part1}Spectral density of the LE-LF (low energy low frequency) $u$-velocity data in Figure~\ref{fig:u-velocity-contours-test41test42test43-part1}, collected at $y^*/\delta=0.6$ and (left to right) $x=604$, 690, 776 and 863. The first row is the narrow bandwidth, continually-excited wavetrain and the second row is the medium bandwidth input pulse. The contours and colors show the relative values of spectral density. Note that the colors are linearly scaled to the maxima (red) and minima (blue) encountered at each location and hence the contours cannot be compared between locations.\\ \centerline{Colorbar: min \includegraphics[height=0.8\baselineskip]{colormap-jet.jpg} max}}
\end{figure}

\begin{figure}
\centering
\includegraphics[width=\linewidth]{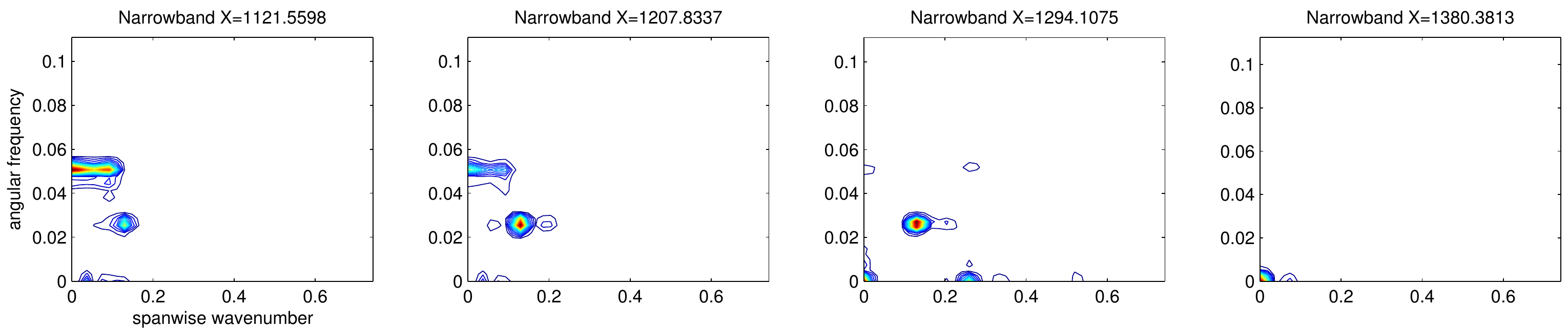}
\includegraphics[width=\linewidth]{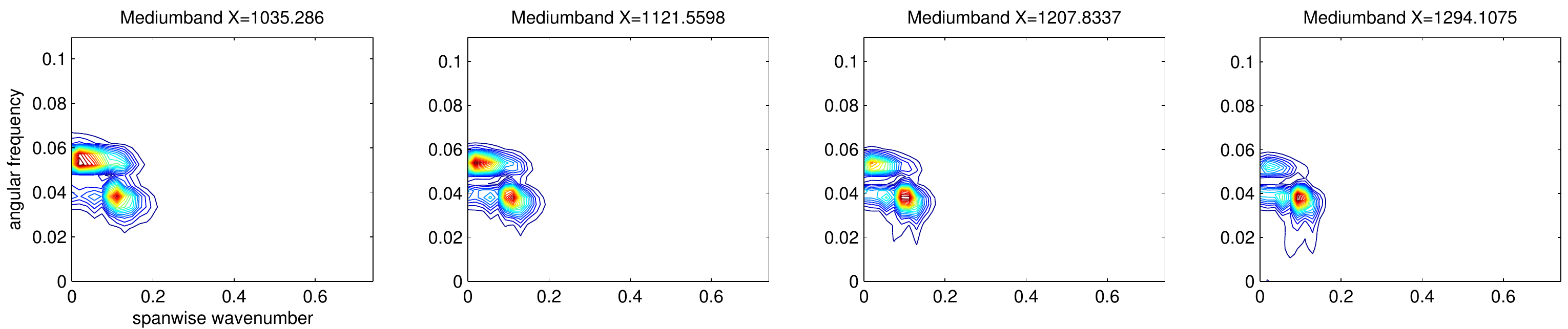}
\caption[]{\label{fig:u-spectra-contours-test41test42test43-part2}Spectral density of the LE-LF (low energy low frequency) $u$-velocity data in Figure~\ref{fig:u-velocity-contours-test41test42test43-part2}, collected at $y^*/\delta=0.6$ and $x$ locations indicated above each plot. The first row is the narrow bandwidth, continually-excited wavetrain and the second row is the medium bandwidth input pulse. The contours and colors show the relative values of spectral density. Note that the colors are linearly scaled to the maxima (red) and minima (blue) encountered at each location and hence the contours cannot be compared between locations.\\ \centerline{Colorbar: min \includegraphics[height=0.8\baselineskip]{colormap-jet.jpg} max}}
\end{figure}

\end{landscape}

We shall focus first on the LE-LF-LB case. According to the linear stability diagram of Figure~\ref{fig:neutral-curve}, the low frequency that forms the main component of the LE-LF-LB wavepacket should be linearly damped as it propagates from the source. However, the sustained excitation provided by the source in time means the perturbation will be present at some reasonable strength even as it decays spatially from the source. The dominant frequency of the wavepacket at $\omega_{max}=0.026$ is apparent as the strongest-energy component of the spectrum at $x=604$ in the first row of Figure~\ref{fig:u-spectra-contours-test41test42test43-part1}. Higher harmonics of the dominant mode have also formed by this time, at $\omega=2(0.026)=0.052$ and $3(0.026)=0.078$. 

\begin{figure}
\centering
\includegraphics{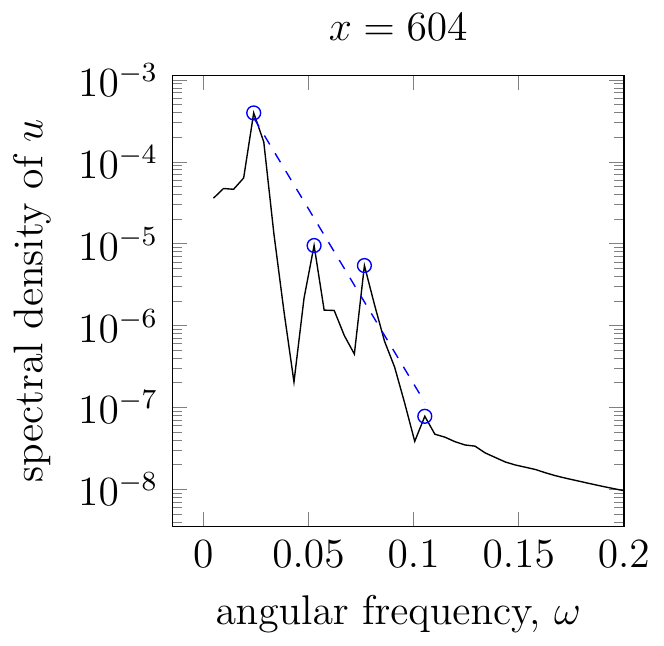}
\caption{\label{fig:LE-LF-LB-1d-geom-progression}One-dimensional spectral density of LE-LF-LB $u$-velocity data collected along the centerline $z=0$ at $x=604$, $y^*/\delta=0.6$. The dashed straight line is the best-fit (in a least-squares sense) geometric progression that connects the harmonic spectral peaks. The common ratio of the geometric progression is $r=0.059$.}
\end{figure}

These harmonics may also be seen in the one-dimensional spectrum of the wavepacket along the centerline at $x=604$ shown in Figure~\ref{fig:LE-LF-LB-1d-geom-progression}, where the attenuation of successively higher harmonics according to an approximately geometric progression is highlighted. Such a sequence of harmonics points to the same phenomena of nonlinear interactions producing multiples of the fundamental frequency as were described by \citet{Kachanov1987}, \citet{Borodulin1995} and \citet{Bake2000}. In \citet{Sengupta2012,Bhaumik2013}, it was also explained how nonlinear growth of the spatio-temporal wave front causes generation of superharmonics of the fundamental frequency (not the excitation frequency) of the wavepacket.

As we progress left to right along the first row of Figure~\ref{fig:u-spectra-contours-test41test42test43-part1}, we find that the second harmonic $(0.0,0.52)$, derived from sum interactions of symmetric oblique wave pairs, amplifies quickly as the wavetrain propagates downstream because is close to the most rapidly growing 2D modes of the boundary layer (see also Figure~\ref{fig:lst-dns-comparison-broadband}a). Meanwhile, we may note that the dominant oblique mode sustained at the source frequency of 0.026 has decayed somewhat, which is consistent with its linearly-damped nature while nonlinear effects remain weak at this early stage. From $x=690$, we may observe a down shifting of its spanwise wavenumber progressing from $(\omega,\beta)\approx(0.026,0.18)$ at $x=690$ to $(0.026,0.14)$ at $x=863$. By $x=863$, we find that the 2D second harmonic of the initial frequency $2\omega_{max}=2(0.026)=0.052$ has grown substantially to become the most dominant mode in the wavepacket. From here, the dominant 2D second harmonic, along with the dominant oblique mode at initial frequency $\omega_{max}= 0.026$, continues to amplify, with the $\beta$-bandwidth of the latter strengthening towards $\beta=0.13$. This progressive down shifting in the $\beta$-bandwidth of the dominant oblique wave modes from $\beta=0.18$ to $\beta\approx 0.13$ is indicative of its parametric interaction with the dominant 2D mode, as $\beta\approx 0.13$ corresponds to the most rapidly growing oblique parametric mode (see \citet{Herbert1984}). The strengthening of the initially damped dominant oblique wave mode may thus be accounted by its parametric amplification via the linearly amplifying 2D mode. By $x=1122$ (first row of Figure~\ref{fig:u-spectra-contours-test41test42test43-part2}) the said modes are in a close Craik triad resonance configuration --- a 2D mode at $(\omega,\beta)\approx(0.052,0)$ and a 3D mode pair at $(\omega,\beta)\approx(0.026,\pm0.13)$ (only the positive $\beta$ portion of the symmetric spectrum is shown). 

Thereafter, the spectrum of $x=1294$ (first row of Figure~\ref{fig:u-spectra-contours-test41test42test43-part2}) shows a system of five-wave resonance, with the near-zero low-frequency mode $(0,0.26)$ forming through nonlinear difference interactions between the subharmonics as $(\omega,\beta)=(0.027,0.13)-(0.027,-0.13)=(0,0.26)$. This new mode has a spanwise wavelength of $\lambda_z=2\pi/\beta=2\pi/0.26=24$, which is very close to the staggered $\Lambda$-vortex spanwise separation in the corresponding $x=1294$ velocity plot in the first row of Figure~\ref{fig:u-velocity-contours-test41test42test43-part2}. The wavepacket is now in the post-subharmonic stage, and the near-zero frequency mode is suggestive of 3D local distortion of the base flow and associated with the rapid growth of the $\Lambda$-vortex system \citep{Yeo2010}. The above points to the overall N-type transition of the wavetrain. 

Furthermore, the above shows that the LE-LF-LB wave resonant triad had formed in the \emph{reverse} sequence from the more typical case. In the wavepackets of \citet{Cohen1991,Medeiros1999b}, the fundamental mode at $\omega_0$ was dominant in the initial wavepacket, and nonlinear difference interactions between component modes described by \citet{Craik2001} seed the subharmonic mode at $\omega_0/2$. In the present low frequency case that is the focus of this section, the mode with subharmonic frequency is excited first at frequency $\omega_0/2=0.026$. The fundamental mode $\omega_0=0.052$ begins at a much smaller amplitude but grows until it is similar in strength to the subharmonic mode. This fundamental mode coincides with the lower branch of the neutral stability curve at the source and thus experiences the longest linear amplification region. Thus what we have appears to be a case of ``delayed'' Craik triad formation, due to the weak 2D fundamental wave of $\omega_0=0.052$. Once the $\omega_0$ and $\omega_0/2$ waves becomes comparable from around $x=863$, the triad takes off, leveraging its growth catalytically upon the growing $\omega_0$ fundamental mode.

In the context of existing knowledge, it is known that the subharmonic mode amplification can be catalyzed by a 2D fundamental wave. The present case shows that even if the fundamental catalyst is initially very weak in the initial wavepacket, it can be amplified slowly until it is of comparable amplitude with the subharmonic. Once the wave triad has formed, regardless of their sequence of origin (subharmonic earlier and fundamental later; or fundamental earlier and subharmonic later), the same mechanism of resonant amplification takes over to cause rapid growth of subharmonic modes leading to staggered $\Lambda$-vortices and N-type transition to turbulence.

Going on now to the medium bandwidth case LE-LF-MB, depicted in the second row of Figures~\ref{fig:u-velocity-contours-test41test42test43-part1} to \ref{fig:u-spectra-contours-test41test42test43-part2}, we note that its wavepacket undergoes a similar transition as its narrow bandwidth counterpart, albeit with some differences. For the start, the shortened duration of the activating source results in substantially weaker oblique waves by virtue of its linearly damped nature. At $x=604$ in the second row of Figure~\ref{fig:u-spectra-contours-test41test42test43-part1}, the frequency bandwidth of the medium bandwidth wavepacket is substantially wider than the earlier narrow bandwidth case. Its amplitude is however only about 1/7 that of the narrow bandwidth case, and its spectral peak is at $(\omega,\beta)=(0.029,0.16)$, which is slightly higher in frequency than the initial disturbance frequency $\omega_{max}=0.026$. 

Higher harmonics are excited at multiples of the initial frequency in the early stages; and a form of spectral triad between a 2D and 3D mode is evident and becomes reasonably distinct by $x=1000-1100$ in the second row of Figure~\ref{fig:u-spectra-contours-test41test42test43-part2}. The disturbance amplitude is weaker than the narrow bandwidth case (LE-LF-LB), and spectrally it never quite achieves the exact triad configuration with dominant subharmonic mode at half the frequency of the fundamental. Instead, as the wavepacket evolves downstream the dominant oblique mode migrates to higher $\omega$ and lower $\beta$, while the strongest 2D mode moves to lower $\omega$. Between $x=604$ (Figure~\ref{fig:u-spectra-contours-test41test42test43-part1}) and $x=1294$ (Figure~\ref{fig:u-spectra-contours-test41test42test43-part2}), the oblique 3D mode shifts from $(\omega,\beta)=(0.029,0.16)$ to $(0.038,0.10)$, while the 2D mode adjusts synchronously from $(\omega,\beta)=(0.058,0)$ to $(0.054,0)$. This has the effect of bringing the 2D and 3D modes closer to each other in spectral space.

Perhaps more importantly, this upward shift of the subharmonic mode shows the greater amplification of positively detuned subharmonic modes. This appears to be supported by a recent study of \citet{Kang2017}, who found that positively detuned subharmonic modes in a Blasius boundary layer could grow faster than tuned Craik resonance in a wide bandwidth wavepacket during the early subharmonic growth stage – we note that the present medium-bandwidth wavepacket has a significant if de-limited frequency bandwidth. This frequency detuning seems also to fit into the framework of \citet{Borodulin2002b} and \citet{Wurz2012a}, who describe asymmetry between the amplification of positively and negatively detuned quasi-subharmonic modes, with larger amplification of the positively detuned modes, although their results strictly speaking pertain to an adverse pressure gradient boundary layer.

Transition of the medium bandwidth wavepacket does not occur within the simulation domain, likely because of the low amplitude of the source disturbance and its shortened excitation duration. The maximum magnitude of the wavepacket u-velocity increases from 0.18\% of the freestream velocity at $x=604$ to 1.12\% at $x=1294$. While the medium bandwidth wavepacket is substantially weaker than the narrow-bandwidth wavepacket, both wavepackets have spectral similarities in their downstream development (Figures~\ref{fig:u-spectra-contours-test41test42test43-part1} and \ref{fig:u-spectra-contours-test41test42test43-part2}) and both their $u$-disturbance contours (Figures~\ref{fig:u-velocity-contours-test41test42test43-part1} and \ref{fig:u-velocity-contours-test41test42test43-part2}) reflects clear signs of staggered formation. 

\begin{figure}
\centering
\includegraphics{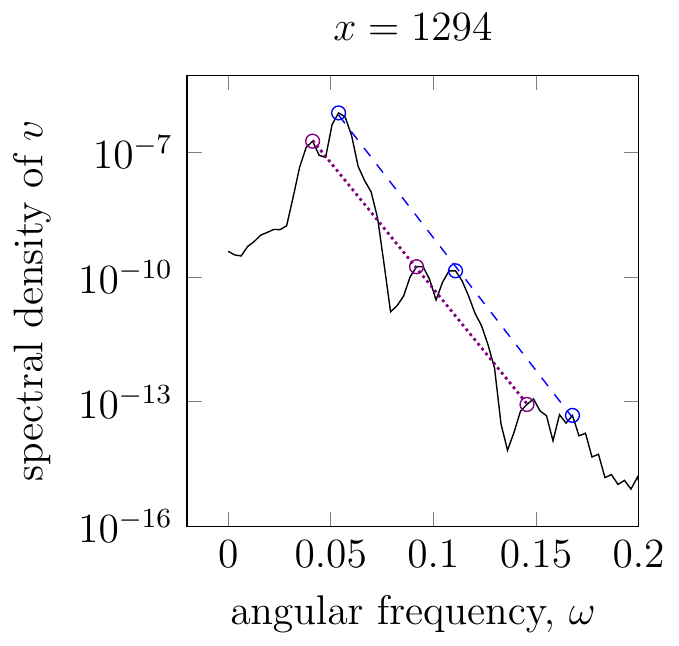}
\caption{\label{fig:LE-LF-MB-1d-geom-progression}One-dimensional spectral density of LE-LF-MB wall-normal $v$-velocity data collected along the centerline $z=0$ at $x=1294$, $y^*/\delta=0.6$. The dashed and dotted straight lines are the best-fit (in a least-squares sense) geometric progressions that connects the harmonic spectral peaks. The common ratio of the geometric progression given by the dashed line is $r_1=2.38\times10^{-4}$ and for the dotted line it is $r_2=8.47\times10^{-4}$.}
\end{figure}

\begin{figure}
\includegraphics{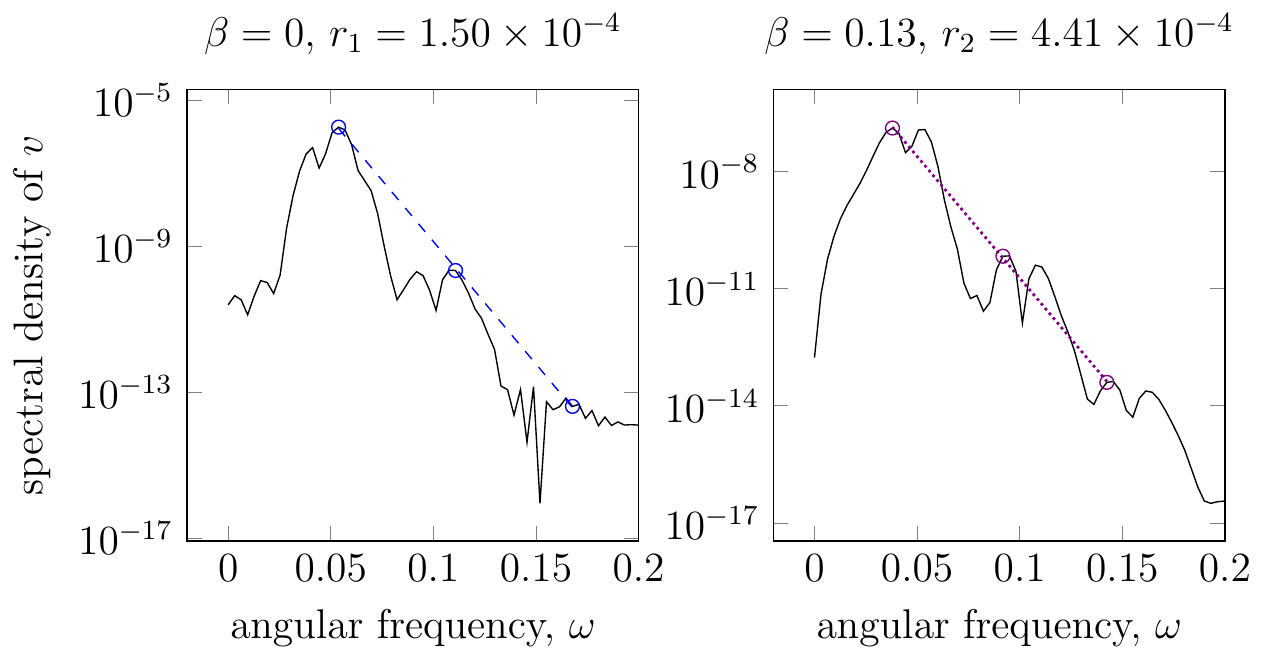}
\caption{\label{fig:LE-LF-MB-2d-geom-progression}Two-dimensional spectral density of LE-LF-MB wall-normal $v$-velocity data collected in a spanwise line (parallel to the $z$-axis) at $x=1294$, $y^*/\delta=0.6$. Line plots extracted from this 2D spectrum at spanwise wavenumber $\beta=0$ and $\beta=0.13$ are plot on the left and right respectively. It can be seen that the harmonic series with fundamental $\omega=0.038$ is dominant for oblique 3D modes $\beta=0.13$, whereas the series with fundamental $\omega=0.054$ is strongest among 2D modes $\beta=0$.}
\end{figure}

Another notable feature of the present medium bandwidth case is the dual series of harmonic peaks obtained from the discrete Fourier transform of wall-normal $v$-velocity data shown in Figure~\ref{fig:LE-LF-MB-1d-geom-progression} ($z=0$, $x=1294$). The lower frequency fundamental peak is at $\omega=0.038$, while the higher frequency fundamental is at $\omega=0.054$, and they are accompanied by their second and third harmonics. A more in-depth analysis of the spectrum was carried out by focusing on waves with $\beta=0$ and $\beta=0.13$ (see Figure~\ref{fig:LE-LF-MB-2d-geom-progression}). By comparing Figure~\ref{fig:LE-LF-MB-1d-geom-progression} with Figure~\ref{fig:LE-LF-MB-2d-geom-progression}, it becomes apparent that the higher-frequency harmonic series containing the $\omega=0.054$ peak is primarily associated with the 2D modes, and the lower-frequency harmonic series with the 3D oblique modes. Nevertheless, the association between the lower-frequency harmonic series and the 3D oblique waves is weaker because the oblique waves ($\beta=0.13$) exist nearly equally both in the lower and in the higher frequency harmonic series.

We also find that in \citet{Sengupta2012,Bhaumik2013} (in the context of 2D flows), for lower amplitude, lower frequency excitation cases, the flow transition is caused by the nonlinear evolution and breakdown of the wavetrain rather than the spatio-temporal wavefront, similar to what is observed in our LE-LF-LB case. It is clear here that for the LE-LF-LB case, the transition is caused by the wavetrain (called the TS wavepacket by \citet{Sengupta2012,Bhaumik2013}) and for the LE-LF-MB case the transition is caused by the spatio-temporal wavefront.

\subsection{\label{sec:mixed-transition} Concurrent N-type and K-type transition}
\begin{figure}
\includegraphics{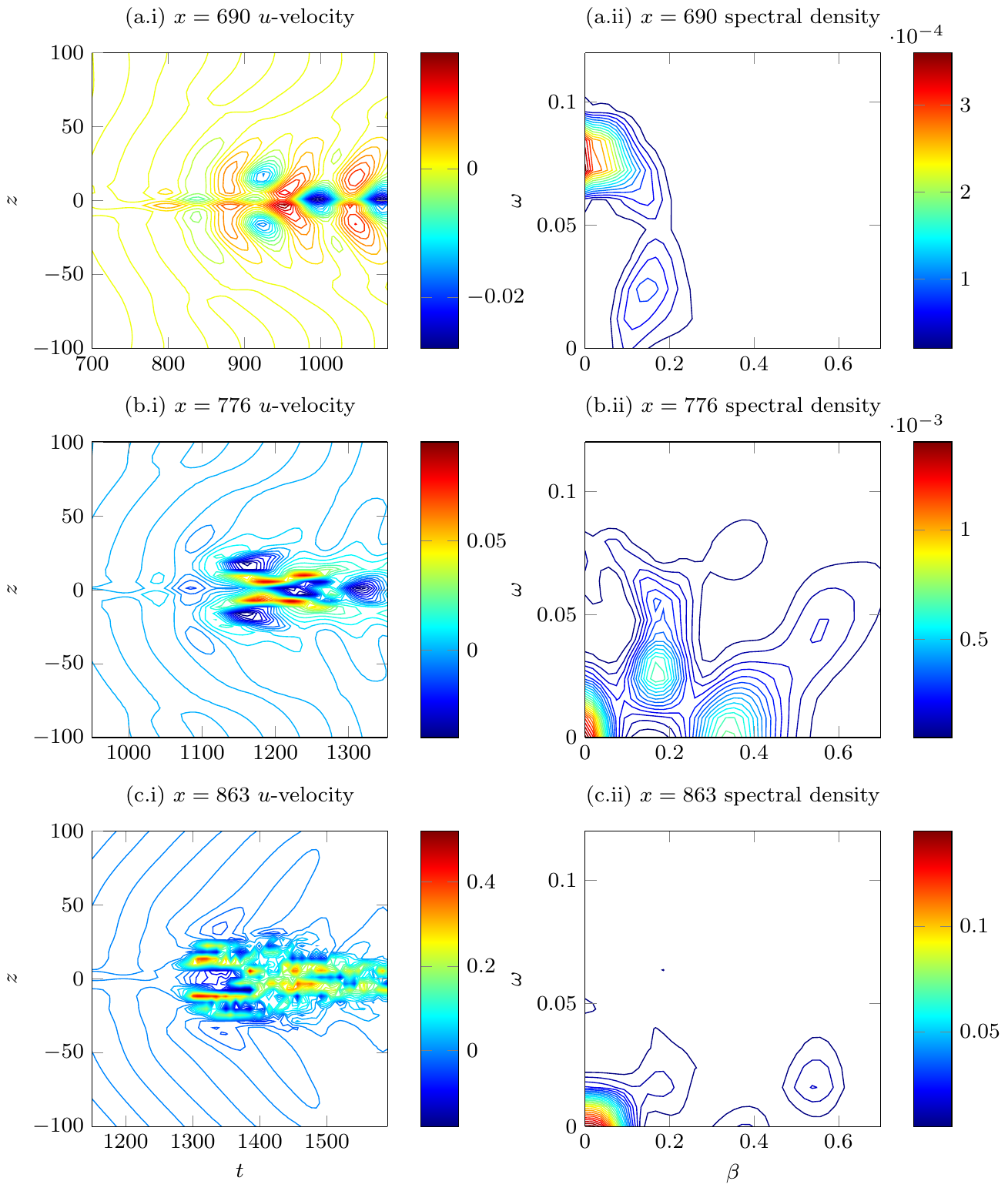}
\caption{\label{fig:u-velocity-spectra-contours-test40s2-wavefront} Velocity contours $u(t,z)$ (left column) and their corresponding spectral density (right column) for the front part of the HE-HF-LB wavetrain only. This wavefront experiences N-type transition. The equivalent velocity contours and spectra for the rest of the wavetrain are shown in Figure~\ref{fig:u-velocity-spectra-contours-test40s2-wavetrainbody}.}
\end{figure}

\begin{figure}
\includegraphics{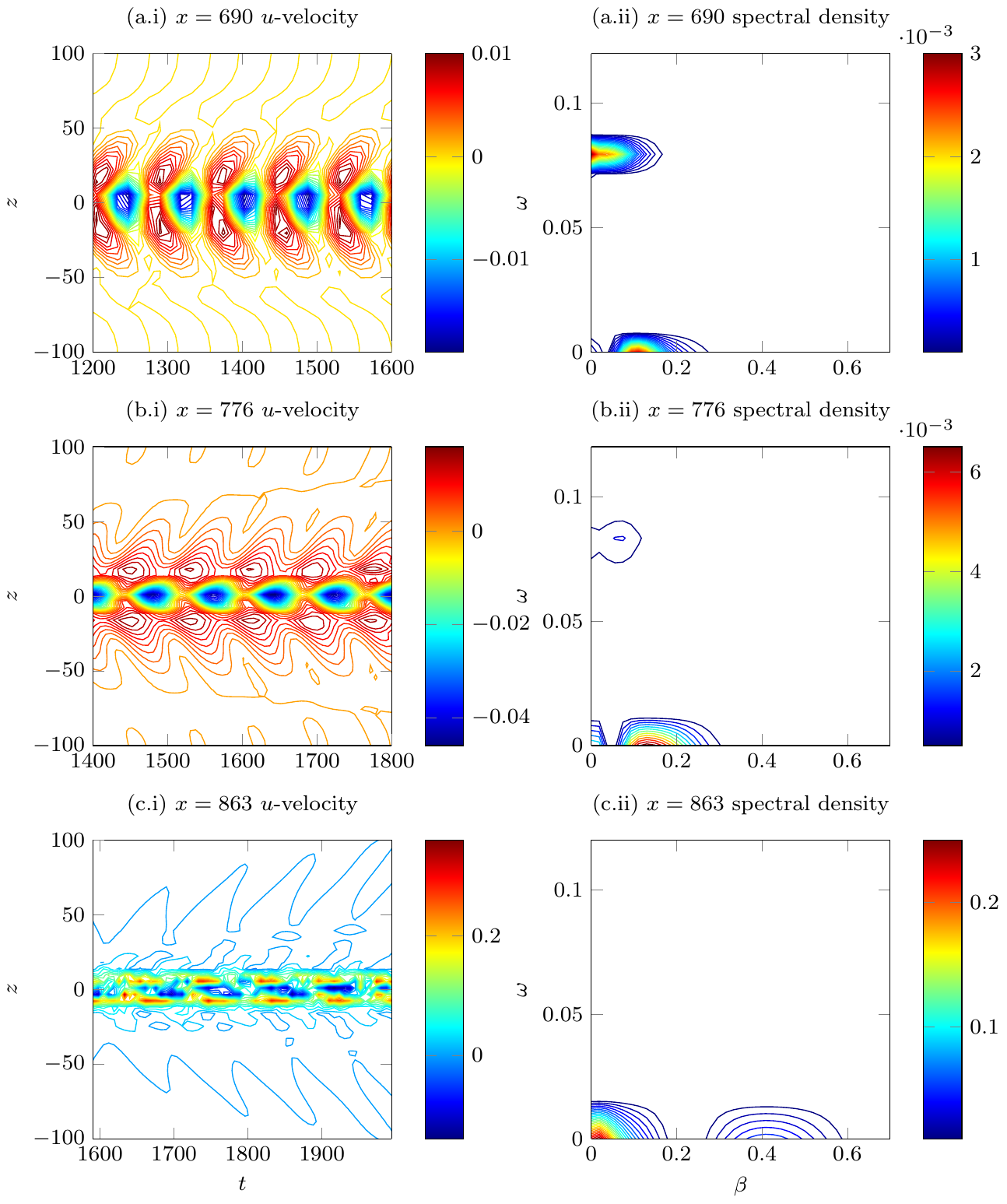}
\caption{\label{fig:u-velocity-spectra-contours-test40s2-wavetrainbody} Velocity contours $u(t,z)$ (left column) and their corresponding spectral density (right column) for the main body of the HE-HF-LB wavetrain. It experiences K-type transition. The equivalent velocity contours and spectra for the wavefront are shown in Figure~\ref{fig:u-velocity-spectra-contours-test40s2-wavefront}.}
\end{figure}

Another interesting phenomenon is presented by the HE-HF-LB case (with high energy, high frequency and narrow bandwidth). Inspection of the velocity contours in the left hand side columns of Figure~\ref{fig:u-velocity-spectra-contours-test40s2-wavefront} and Figure~\ref{fig:u-velocity-spectra-contours-test40s2-wavetrainbody} show that the front and main body of the wavetrain undergo different transition behaviour.

A plausible explanation for this is that the frontal edge of the wavetrain is a form of streamwise inhomogeneity, which represents in the course of its propagation a form of spatial-temporal modulation acting on the leading wave \citep{Healey1995}. This causes the wave front to behave like a part of a broad or medium bandwidth wavepacket (especially at high frequency and/or high energy); thus containing within this limited section of the wavetrain potential seeds for a subharmonic mode, and setting the stage for a N-type transition \citep{Cohen1991, Medeiros1999a, Medeiros1999b, Yeo2010}. The vulnerable region is, however, largely confined to the front of the wavetrain and it should not be considered as part of a truly narrowband disturbance. The main body of the wavetrain (the ``true'' narrowband disturbance) is primarily susceptible to K-type transition \citep{Medeiros2004}. 

To shed further light on the above scenario, discrete Fourier transforms were carried out separately on the wavefront and the main body of the HE-HF-LB wavetrain. The results are illustrated in  in the right hand side columns of Figure~\ref{fig:u-velocity-spectra-contours-test40s2-wavefront} and Figure~\ref{fig:u-velocity-spectra-contours-test40s2-wavetrainbody}. At $x=690$ in Figure~\ref{fig:u-velocity-spectra-contours-test40s2-wavefront}(a.ii), it can be seen that the frontal section of the wavetrain comprises a strong broadband 2D fundamental mode at $\omega=0.08$ and a weaker quasi-subharmonic mode at $(\omega,\beta)=(0.024,0.15)$, while the main wavetrain in Figure~\ref{fig:u-velocity-spectra-contours-test40s2-wavetrainbody}(a.ii) has a narrow band 2D mode at the same $\omega=0.08$ and a co-dominant low-frequency oblique mode at $(\omega,\beta)=(0,0.1)$.The velocity contours in Figure~\ref{fig:u-velocity-spectra-contours-test40s2-wavetrainbody}(a.i) reveal that the spanwise structures of main body consists of one large velocity perturbation in the center flanked by two perturbations of an opposite sign at the side (such as one negative central region flanked by two positive side lobes), meaning that it is roughly covering 1.5 cycles of a spanwise periodic disturbance. With the wavetrain spanwise width of about 100, this implies a spanwise wavelength $\lambda_z=100/1.5\approx67$ and spanwise wavenumber $2\pi/67=0.09$, which is very close to the spectral peak of the oblique wave modes at $(\omega,\beta)=(0,0.1)$ given in Figure~\ref{fig:u-velocity-spectra-contours-test40s2-wavetrainbody}(a.ii).

By $x=776$, Figure~\ref{fig:u-velocity-spectra-contours-test40s2-wavefront}(b.i) shows that the wavefront region has developed a horseshoe vortex type of structure, with a resonating system of modes and N-type transition in Figure~\ref{fig:u-velocity-spectra-contours-test40s2-wavefront}(b.ii). On the other hand, no triad resonance is observable in the spectrum at $x=776$ for the main body of the wavetrain in Figure~\ref{fig:u-velocity-spectra-contours-test40s2-wavetrainbody}(b.ii), whose transition is clearly of K-type, with streaks or Klebanoff modes becoming visible at $x=863$ in Figure~\ref{fig:u-velocity-spectra-contours-test40s2-wavetrainbody}(c.i). These streaks have a spacing in the $z$-direction of around $15$ units, corresponding to a spanwise wavenumber $2\pi/15=0.42$, identifying it with the $\beta\approx0.4$ low-frequency peak in the spectrum of Figure~\ref{fig:u-velocity-spectra-contours-test40s2-wavetrainbody}(c.ii).

\begin{figure}
\includegraphics{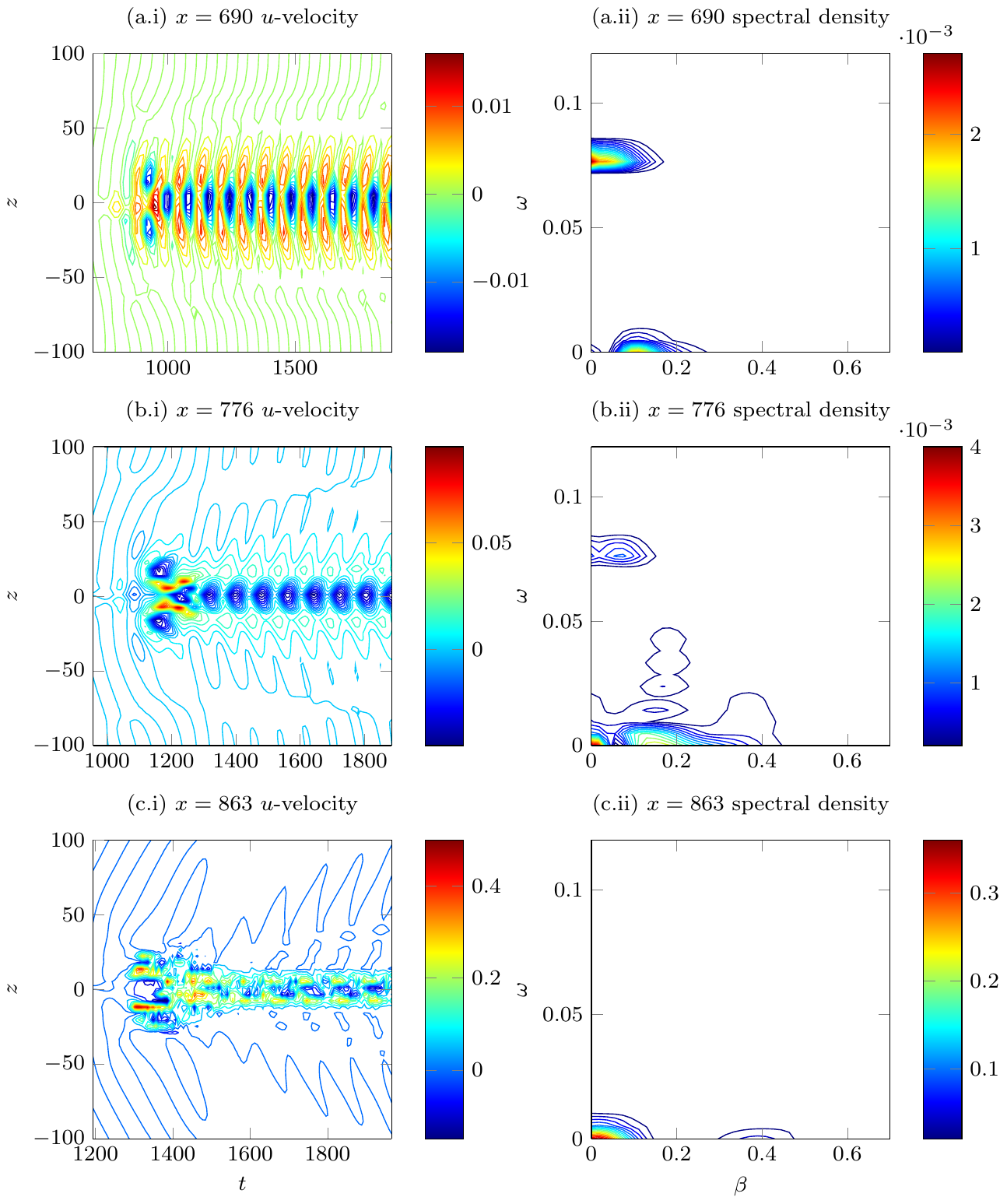}
\caption{\label{fig:u-velocity-spectra-contours-test40s2-full} Velocity contours $u(t,z)$ (left column) and their corresponding spectral density (right column) for the full HE-HF-LB wavetrain (wavefront and main body together). This figure should be compared with the results for the localized wavefront and main wavetrain body in Figure~\ref{fig:u-velocity-spectra-contours-test40s2-wavefront} and Figure~\ref{fig:u-velocity-spectra-contours-test40s2-wavetrainbody} respectively.}
\end{figure}

Comparing these separate spectra for the wavefront region and main body of the wavetrain with the spectrum of the total wavetrain in Figure~\ref{fig:u-velocity-spectra-contours-test40s2-full}, we find that the total spectrum bears much closer resemblance to the spectrum of the main body of the wavetrain. Since the spectral density is a measure of the energy contained in the modes (via the Plancherel theorem), it shows that for the sample lengths we use, the energy contained in the main body of the wavetrain undergoing K-type transition is much greater than the energy in the modulated frontal section experiencing N-type transition. An observer with access to the total spectrum would naturally and rightly conclude that the wavetrain has undergone a K-type transition. 

One may further remark that wavefront/edge modulation tends to be accentuated by high disturbance energy and/or high frequency. On the other hand, where the present broad and medium bandwidth wavepackets are concerned, their small spatial-temporal range renders it difficult to distinguish a distinctive core that is clearly isolated from the edges. A global or total characterization of the entire wavepacket is sufficient and separate analysis of the edge effects seems unnecessary or may not be feasible.

\subsection{\label{sec:peak-freq-jump} Continuous and abrupt peak frequency shift}
\begin{figure}
\includegraphics{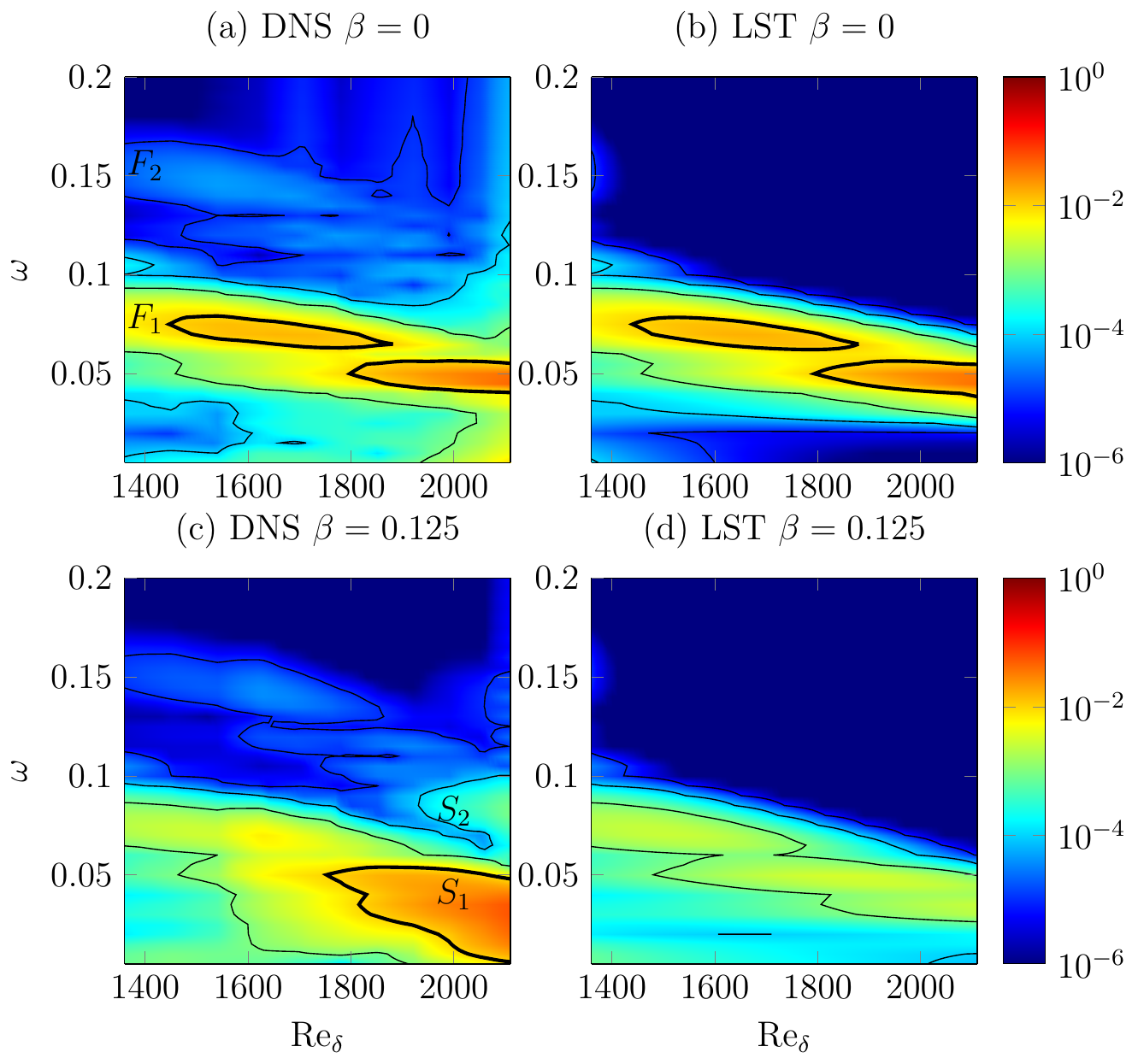}
\caption{\label{fig:test39sine}Frequency spectrum development of case LE-HF-MB. The black contour lines are on a base-10 logarithmic scale, such that successive contour levels are $\lbrace10^{-5},10^{-4},10^{-3},...\rbrace$, with the contour line for $10^{-2}$ thicker than the rest. Additional detail is provided through a background shaded according to the colorbar shown. The initial fundamental and its harmonic are labelled $F_1$ and $F_2$ in (a). The approximately corresponding subharmonic modes are labelled $S_1$ and $S_2$ in (c).}
\end{figure}

A case that exhibits an abrupt shift in the dominant frequency (maximum spectral density, $\omega_{max}$) is perhaps the LE-HF-MB (low energy, high frequency, medium bandwidth) wavepacket, whose frequency spectra is plotted as function of displacement thickness Reynolds number $\mathrm{Re}_\delta$ in Figure~\ref{fig:test39sine}. Here the dominant 2D frequency $F_1$ in Figure~\ref{fig:test39sine}a undergoes a fairly abrupt shift from $\omega\approx0.07$ to $\omega\approx0.05$ at $\mathrm{Re}_\delta\approx1800$. 

We find that this is part of a trend across many of the wavepackets investigated - they show a shift in their dominant frequency as convection downstream occurs. $\omega_{max}$ is found to tend towards $\omega=0.0528$, which is the frequency of the lower branch of the neutral stability curve at the disturbance source. In other words, $\omega_{max}$ refers to the most energetic frequency present in the wavepacket at one streamwise location; when the wavepacket spectrum is compared at different streamwise locations, we observe a common trend in the decrease of $\omega_{max}$ as the wavepacket convects downstream, until $\omega_{max}\approx0.0528$. In most cases, the shift in frequency is continuous, gradual, and may be explained on the basis of linear stability theory (LST). The abrupt shift in frequency of LE-HF-MB can also be explained by LST, however, when we note that the $\omega\approx0.05$ wave mode which starts at branch I of the neutral curve could continue to grow for the longest spatial distance, even as nearby modes such as $\omega\approx0.07$ go past their growth peaks and travel into slower growth or linearly damped Reynolds number regimes. 

This continuous shift in the frequency content of the wavepacket towards lower values during its downstream propagation, while showing a sudden jump during its later evolution stages, is noted in \citet{Sengupta2012, Bhaumik2013}. Moreover, similar LST-supported features of mode competition have been recently observed in DNS of wavepackets in a supersonic compression corner flow. The station of the disturbance source dictated the winning mode because of different downstream growth properties for different modes, in particular, different lower branches of neutral curves. Details can be found in \citet{Novikov2016}.

The second harmonic of the fundamental, $F_2$ is not predictable by LST, and is hence absent from Figure~\ref{fig:test39sine}b. 
Nevertheless, the presence of $F_2$ is supported by the $F_1$ mode through nonlinear mechanisms and $F_2$ may be seen to broadly track the fundamental $F_1$ in Figure~\ref{fig:test39sine}a. In addition, there is also evidence of corresponding dominant subharmonic frequency shifting to track the observed change in the dominant fundamental frequency in the DNS plot of $\beta=0.125$ modes in Figure~\ref{fig:test39sine}c, where the subharmonic band labelled $S_1$ shifts to lower frequencies as $Re_\delta$ increases. A second harmonic of the subharmonic mode also seems to be present, and it is marked as $S_2$.

\begin{figure}
\includegraphics{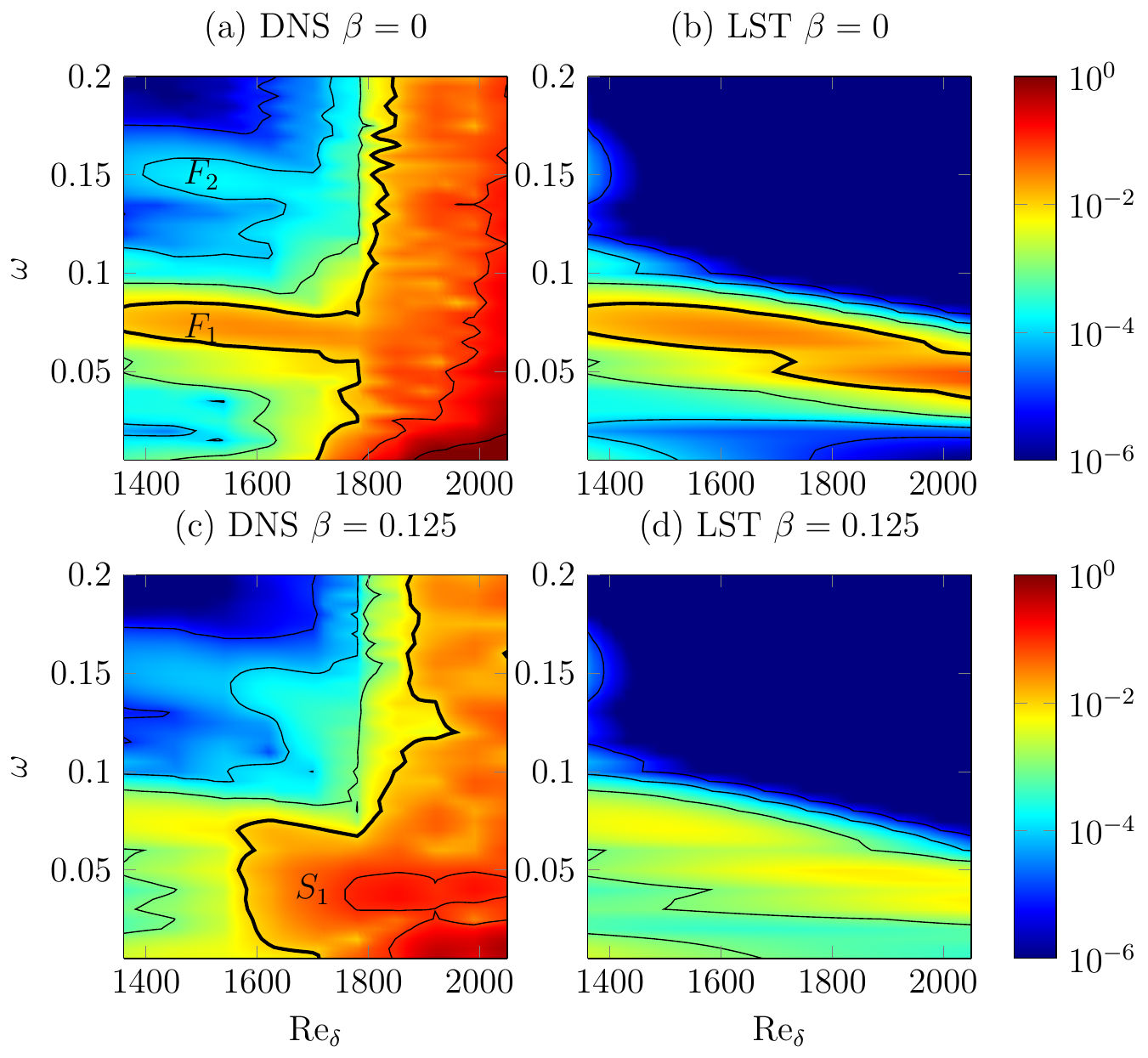}
\caption{\label{fig:test39sine2}Frequency spectrum evolution of case HE-HF-MB. The black contour lines are on a base-10 logarithmic scale, such that successive contour levels are $\lbrace10^{-5},10^{-4},10^{-3},...\rbrace$, with the contour line for $10^{-2}$ thicker than the rest. Additional detail is provided through a background shaded according to the colorbar shown.}
\end{figure}

On the other hand, if the initial amplitude of the disturbance is sufficiently large (HE-), a resonant triad at the higher fundamental 2D frequency of the source may evolve to dominate the transition process, and the shift of the 2D fundamental to the most linearly dominant 2D frequency seen earlier for the lower energy (LE-) case is effectively bypassed. This is exemplified by the more energetic HE-HF-MB (high energy, high frequency, medium bandwidth) case in Figure~\ref{fig:test39sine2}. Here, rapid and abrupt expansion of the 2D and $\beta=0.125$ DNS spectra towards high frequency around $Re_\delta\approx 1800$ signals its accelerated breakdown to incipient turbulence relative to the preceding lower energy case. 

Another way to put this is that high energy input to higher frequency modes could favor them to reach the threshold for nonlinear breakdown earlier, hence reducing their reliance on the linear mechanism of disturbance growth – a form of bypass transition.

\subsection{\label{sec:transition-types}Summary of transition scenarios}
In this work, a few cases with special or unique transition behavior were observed and discussed. These are summarized below. A summary about all the 18 wavepackets simulated in a full factorial design is then given in Table~\ref{tab:summary}, while further details about them are available in \citet{Kang2015a}. The special cases that were elaborated upon are:
\begin{description}
\item[Reverse triad formation sequence] \hfill \\
The typical wavepacket begins with modes in a fundamental 2D wave band around $\omega_0$, and gradually develops a subharmonic mode $\omega_0/2$ that begins a process of resonance with the fundamental mode in a Craik-type resonant triad. Our simulations have revealed the reverse is also possible - the development of a resonant triad from a wavepacket whose initial dominant frequency is $\omega_0/2$. The $\omega_0$ mode can grow linearly from a very small initial amplitude until it is of sufficient size to resonate with $\omega_0/2$. This process is exemplified by the LE-LF-MB (low energy, low frequency, medium bandwidth) and LE-LF-LB (low energy, low frequency, narrow bandwidth) cases, which were discussed in Section~\ref{sec:reverse-triad-sequence}. This variant of the triad formation process is believed to be new and has not been discussed in the literature before.

\item[Concurrent N-type and K-type transition] \hfill \\
In some of the long wavepackets (wavetrains) that we study, different transition behavior is observed at the front and in the main body of the wavetrain. When we find such local modulation \citep{Healey1995} or edge effects to be present, we clarify the picture by performing additional Fourier analysis on each part of the wavetrain separately, such as is done for case HE-HF-LB (high energy, high frequency, narrow bandwidth) in Section~\ref{sec:mixed-transition}.

\item[Continuous and abrupt peak frequency shift] \hfill \\
Both the linear stability theory (LST) and DNS computations show the existence of a ``preferred'' wavepacket frequency. Wavepackets that are initially dominated by waves of higher frequencies display a tendency to shift their dominant frequency towards the preferred frequency, which corresponds to the 2D ($\beta=0$) frequency of the lower branch of the neutral stability curve at the disturbance source. This preference is ``overridden'' by increasing the amplitude of the dominant frequency. The frequency peak shift may occur in a fairly abrupt manner; which is believed to be related to large differences in linear growth and decay rates between waves. We explored these in Section~\ref{sec:peak-freq-jump} for the LE-HF-MB (low energy, high frequency, medium bandwidth) and HE-HF-MB (high energy, high frequency, medium bandwidth) wavepackets.
 
\end{description}

Table~\ref{tab:summary} provides a summary of the transition behavior for the other wavepackets investigated in this factorial study. In most cases, breakdown to incipient turbulence occurs within the simulation domain. Nevertheless, there are several cases in which breakdown does not occur within the domain, and these cases are denoted by the $^{\#}$ symbol against the anticipated route of transition. There are also some wavetrains with N-type transition in front and K-type in their main body, marked as N\textsuperscript{F}K\textsuperscript{B}. Here, the superscripts ``F'' and ``B'' refer to the front and main body of the wavetrain respectively.

\begin{table}[h]
\caption{\label{tab:summary}Summary of the transition type for each case. K denotes K-type transition with streaky structures and aligned $\Lambda$-vortices, while N denotes subharmonic N-type transition with staggered $\Lambda$-vortices. The superscripts F and B indicate that the transition type applies to the transient front and the main body of the wavetrain respectively. The $^{\#}$ symbol indicates no breakdown to turbulence within the domain.}
\begin{tabular*}{\hsize}{@{\extracolsep{\fill}}llll@{}}
\hline
\textbf{LOW ENERGY} & \textbf{3dB Bandwidth ($\Delta\omega$)} &  &\\
\textbf{Angular Frequency ($\omega_{max}$)} & Low & Medium & High\\
\hline
Low & N & N$^{\#}$ & N\\
Medium & K & N & N\\
High & N\textsuperscript{F}K\textsuperscript{B} & N$^{\#}$ & N$^{\#}$ \\
\hline
\end{tabular*}

\begin{tabular*}{\hsize}{@{\extracolsep{\fill}}llll@{}}
\hline
\textbf{HIGH ENERGY} & \textbf{3dB Bandwidth ($\Delta\omega$)} &  &\\
\textbf{Angular Frequency ($\omega_{max}$)} & Low & Medium & High\\
\hline
Low & K & N & N\\
Medium & K & K & N\\
High & N\textsuperscript{F}K\textsuperscript{B} & N & N\\
\hline
\end{tabular*}
\end{table}

From Table~\ref{tab:summary} one may observe the following trends:
\begin{enumerate}
\item Input disturbances that are wide in terms of frequency bandwidth unequivocally encourage N-type transition. This may be seen from the (right-most) column 4 of Table~\ref{tab:summary}, where all wide bandwidth wavepackets undergo N-type transition. 
\item As the frequency bandwidth narrows (from column 4 to 2) instances of K-type transition begin to emerge so that K-type transition largely predominates for narrow bandwidth disturbances - an exception applies for low-energy low-frequency (LE-LF-) disturbances which could still adhere to an N-type transition. Interestingly, at high frequency, the front of the wavetrain could exhibit N-type characteristics while the main body of the wave undergoes K-type transition, as investigated in Section~\ref{sec:mixed-transition}. K-type transition for higher-frequency disturbances had also been reported by \citet{Bhaumik2014a}.
\item Wavepackets arising from low energy disturbances tend to experience mostly N-type transition – although narrowing the bandwidth results in an emerging K-type mechanism. Increasing the disturbance energy levels encourages more medium and narrow bandwidth disturbances to switch over to K-type transition, while high bandwidth disturbances nevertheless continue to follow the N-type route.
\item Where intermediate bandwidth (-MB) disturbances are concerned (column 3) there is a preponderant tendency to follow the N-type transitional route, although increase in the energy level may encourage some to follow the K-type route.
\end{enumerate}

\section{\label{sec:conc}Conclusions}
In conclusion, not all wavepackets encountered in nature are the same. The present work represents in a sense a more systematic and in-depth examination of the process of wavepacket transition, where a range of key wavepacket properties were studied to understand how they might potentially influence the transitional route or pathway a wavepacket may follow. The spatial-temporal evolution of the wavepackets activated by a point-like source was modelled by direct numerical simulation (DNS) in the setting of a Blasius boundary layer. Eighteen different combinations of wavepacket frequency, amplitude and bandwidth were studied in terms of their disturbance structures and spectral properties to elucidate the major trends and exceptions. It was often found that broad and narrow bandwidth wavepackets led to N-type and K-type transition respectively.

Within this general classification scheme of N-type and K-type transition, many variants in the details of the transition process are found to occur. Three of these were highlighted in this paper.

The first is a novel transition path variant of the Craik-type resonant triads for low-energy wavepackets with low frequency in a narrow band that may be termed a ``reverse Craik triad formation sequence''. The typical Craik resonant triads begin with a pre-existing fundamental mode, and a subharmonic mode that develops upon it. For an initial wavepacket with frequency $\omega_{max}$ that is half of the preferred fundamental frequency, a reverse formation may take place first with the gradual growth of the linearly preferred $2\omega_{max}$ mode. The linearly amplifying $2\omega_{max}$ mode then acts as the 2D fundamental to the original $\omega_{max}$ oblique waves in a Craik-type triad that dominates the transition to breakdown. A variation to this may happen for wider-band low frequency wavepackets. In this case, the wider bandwidth permits the selection of a positively-detuned subharmonic triad as the leading driver of wavepacket growth. The predominance of a positively-detuned subharmonic for broadband wavepackets in a Blasius boundary layer has been noted by \citet{Kang2017} recently, though its importance in adverse pressure gradient boundary layers had already been noted in the experiments of \citep{Wurz2012a}.

The second observation pertains to narrow-band wavetrains, whose long streamwise length allows them to transition differently at the front and in the main body of the wavetrain. In such instances, the wavefront propagates into a largely quiescent, unperturbed boundary layer and thus experiences localized spatial-temporal variations that act like a broadband source that favors N-type processes. The narrow-band homogeneity of the main core, however, limits the rearward infiltration of the wavefront perturbations. The final result is a wavepacket that undergoes K-type transition in the main body of the wavepacket, while a limited frontal region experiences N-type breakdown. The dominance of the extended main body means that this type of hybrid transition may not be discernible from the spectral analysis of the whole wavepacket, and would only show up via a narrow-windowed Fourier analysis. The mixed-mode transition tends to be encouraged by high source frequency and/or high energy, since these tend to accentuate frontal inhomogeneity.  

The third observation is that boundary transition is strongly underpinned by linear stability. Where an initiated packet of wave disturbance deviates from the linearly preferred growth path, it has a tendency in the course of its evolution to shift or drift back into the nearest linearly favored path, as the most favored modes thrive and others may decay. This frequently accounts for the gradual shift in the dominant 2D frequencies observed during the transition, although sometimes the drift may be more abrupt or rapid. This essentially classical scenario may be upset, however, when the energy of the initiated wavepacket is high, so that nonlinear processes dominate the transition before linear mechanisms could assert their role. 



\bibliographystyle{elsarticle-harv} 
\bibliography{references}

\begin{thebibliography}{54}
\expandafter\ifx\csname natexlab\endcsname\relax\def\natexlab#1{#1}\fi
\providecommand{\url}[1]{\texttt{#1}}
\providecommand{\href}[2]{#2}
\providecommand{\path}[1]{#1}
\providecommand{\DOIprefix}{doi:}
\providecommand{\ArXivprefix}{arXiv:}
\providecommand{\URLprefix}{URL: }
\providecommand{\Pubmedprefix}{pmid:}
\providecommand{\doi}[1]{\href{http://dx.doi.org/#1}{\path{#1}}}
\providecommand{\Pubmed}[1]{\href{pmid:#1}{\path{#1}}}
\providecommand{\bibinfo}[2]{#2}
\ifx\xfnm\relax \def\xfnm[#1]{\unskip,\space#1}\fi
\bibitem[{Bake et~al.(2000)Bake, Fernholz and Kachanov}]{Bake2000}
\bibinfo{author}{Bake, S.}, \bibinfo{author}{Fernholz, H.H.},
  \bibinfo{author}{Kachanov, Y.S.}, \bibinfo{year}{2000}.
\newblock \bibinfo{title}{Resemblance of {K-} and {N-regimes} of boundary-layer
  transition at late stages}.
\newblock \bibinfo{journal}{European Journal of Mechanics - B/Fluids}
  \bibinfo{volume}{19}, \bibinfo{pages}{1--22}.
\newblock \URLprefix
  \url{http://www.sciencedirect.com/science/article/pii/S0997754600001084},
  \DOIprefix\doi{http://dx.doi.org/10.1016/s0997-7546(00)00108-4}.
\bibitem[{Bhaumik(2013)}]{Bhaumik2013}
\bibinfo{author}{Bhaumik, S.}, \bibinfo{year}{2013}.
\newblock \bibinfo{title}{Direct numerical simulation of inhomogeneous
  transitional and turbulent flows}.
\newblock Ph.D. thesis. IIT Kanpur.
\bibitem[{Bhaumik and Sengupta(2014)}]{Bhaumik2014a}
\bibinfo{author}{Bhaumik, S.}, \bibinfo{author}{Sengupta, T.K.},
  \bibinfo{year}{2014}.
\newblock \bibinfo{title}{Precursor of transition to turbulence: Spatiotemporal
  wave front}.
\newblock \bibinfo{journal}{Phys. Rev. E} \bibinfo{volume}{89},
  \bibinfo{pages}{043018}.
\newblock \URLprefix \url{http://dx.doi.org/10.1103/PhysRevE.89.043018}.
\bibitem[{Bhaumik and Sengupta(2015)}]{Bhaumik2015}
\bibinfo{author}{Bhaumik, S.}, \bibinfo{author}{Sengupta, T.K.},
  \bibinfo{year}{2015}.
\newblock \bibinfo{title}{A new velocity–vorticity formulation for direct
  numerical simulation of 3d transitional and turbulent flows}.
\newblock \bibinfo{journal}{J. Comput. Phys.} \bibinfo{volume}{284},
  \bibinfo{pages}{230 -- 260}.
\newblock \URLprefix
  \url{http://www.sciencedirect.com/science/article/pii/S002199911400847X},
  \DOIprefix\doi{https://doi.org/10.1016/j.jcp.2014.12.030}.
\bibitem[{Bhaumik et~al.(2015)Bhaumik, Sengupta, Mudkavi and
  Mulloth}]{Bhaumik2015a}
\bibinfo{author}{Bhaumik, S.}, \bibinfo{author}{Sengupta, T.K.},
  \bibinfo{author}{Mudkavi, V.}, \bibinfo{author}{Mulloth, A.},
  \bibinfo{year}{2015}.
\newblock \bibinfo{title}{Different routes of transition by spatio-temporal
  wave-front}, in: \bibinfo{editor}{Sengupta, T.K.}, \bibinfo{editor}{Lele,
  S.K.}, \bibinfo{editor}{Sreenivasan, K.R.}, \bibinfo{editor}{Davidson, P.A.}
  (Eds.), \bibinfo{booktitle}{Advances in Computation Modeling and Control of
  Transitional and Turbulent Flows}, \bibinfo{publisher}{World Scientific
  Publishing Company}. pp. \bibinfo{pages}{68--83}.
\bibitem[{Bhaumik et~al.(2017)Bhaumik, Sengupta and Shabab}]{Bhaumik2017a}
\bibinfo{author}{Bhaumik, S.}, \bibinfo{author}{Sengupta, T.K.},
  \bibinfo{author}{Shabab, Z.A.}, \bibinfo{year}{2017}.
\newblock \bibinfo{title}{Receptivity to harmonic excitation following
  nonimpulsive start for boundary-layer flows}.
\newblock \bibinfo{journal}{AIAA J.} \bibinfo{volume}{55},
  \bibinfo{pages}{3233--3238}.
\newblock \URLprefix \url{https://doi.org/10.2514/1.J056292},
  \DOIprefix\doi{10.2514/1.J056292},
  \href{http://arxiv.org/abs/https://doi.org/10.2514/1.J056292}{{\tt
  arXiv:https://doi.org/10.2514/1.J056292}}.
\bibitem[{Birkhoff et~al.(1962)Birkhoff, Varga and Young}]{Birkhoff1962}
\bibinfo{author}{Birkhoff, G.}, \bibinfo{author}{Varga, R.S.},
  \bibinfo{author}{Young, D.}, \bibinfo{year}{1962}.
\newblock \bibinfo{title}{Alternating Direction Implicit Methods}.
  \bibinfo{publisher}{Elsevier}. volume~\bibinfo{volume}{3}.
\newblock pp. \bibinfo{pages}{189--273}.
\newblock \URLprefix
  \url{http://www.sciencedirect.com/science/article/pii/S0065245808606208},
  \DOIprefix\doi{http://dx.doi.org/10.1016/S0065-2458(08)60620-8}.
\bibitem[{Boiko et~al.(2012)Boiko, Dovgal, Grek and Kozlov}]{Boiko2012}
\bibinfo{author}{Boiko, A.V.}, \bibinfo{author}{Dovgal, A.V.},
  \bibinfo{author}{Grek, G.R.}, \bibinfo{author}{Kozlov, V.V.},
  \bibinfo{year}{2012}.
\newblock \bibinfo{title}{Physics of Transitional Shear Flows}.
  \bibinfo{publisher}{Springer}. chapter~\bibinfo{chapter}{12}.
\newblock Number~\bibinfo{number}{98} in \bibinfo{series}{Fluid Mechanics and
  Its Applications}, pp. \bibinfo{pages}{225--226}.
\bibitem[{Borodulin and Kachanov(1995)}]{Borodulin1995}
\bibinfo{author}{Borodulin, V.I.}, \bibinfo{author}{Kachanov, Y.S.},
  \bibinfo{year}{1995}.
\newblock \bibinfo{title}{Formation and development of coherent structures in a
  transitional boundary layer}.
\newblock \bibinfo{journal}{J. Appl. Mech. Tech. Phys.} \bibinfo{volume}{36},
  \bibinfo{pages}{532--564}.
\newblock \URLprefix \url{http://dx.doi.org/10.1007/bf02371277}.
\bibitem[{Borodulin et~al.(2002)Borodulin, Kachanov, Koptsev and
  Roschektayev}]{Borodulin2002b}
\bibinfo{author}{Borodulin, V.I.}, \bibinfo{author}{Kachanov, Y.S.},
  \bibinfo{author}{Koptsev, D.B.}, \bibinfo{author}{Roschektayev, A.P.},
  \bibinfo{year}{2002}.
\newblock \bibinfo{title}{Experimental study of resonant interactions of
  instability waves in self-similar boundary layer with an adverse pressure
  gradient: {II}. {Detuned} resonances}.
\newblock \bibinfo{journal}{J. Turbul.} \bibinfo{volume}{3},
  \bibinfo{pages}{1--22}.
\newblock \URLprefix \url{http://dx.doi.org/10.1088/1468-5248/3/1/063}.
\bibitem[{Brandt(1977)}]{Brandt1977}
\bibinfo{author}{Brandt, A.}, \bibinfo{year}{1977}.
\newblock \bibinfo{title}{Multi-level adaptive solutions to boundary-value
  problems}.
\newblock \bibinfo{journal}{Math. Comp.} \bibinfo{volume}{31},
  \bibinfo{pages}{333--390}.
\newblock \URLprefix \url{http://dx.doi.org/10.1090/S0025-5718-1977-0431719-X}.
\bibitem[{Chorin(1969)}]{Chorin1969}
\bibinfo{author}{Chorin, A.J.}, \bibinfo{year}{1969}.
\newblock \bibinfo{title}{On the convergence of discrete approximations to the
  {Navier-Stokes} equations}.
\newblock \bibinfo{journal}{Math. Comp.} \bibinfo{volume}{23},
  \bibinfo{pages}{341--353}.
\newblock \URLprefix \url{http://dx.doi.org/10.1090/S0025-5718-1969-0242393-5}.
\bibitem[{Cohen(1994)}]{Cohen1994}
\bibinfo{author}{Cohen, J.}, \bibinfo{year}{1994}.
\newblock \bibinfo{title}{The initial evolution of a wave packet in a laminar
  boundary layer}.
\newblock \bibinfo{journal}{Phys. Fluids} \bibinfo{volume}{6},
  \bibinfo{pages}{1133--1143}.
\newblock \URLprefix
  \url{http://scitation.aip.org/content/aip/journal/pof2/6/3/10.1063/1.868284},
  \DOIprefix\doi{http://dx.doi.org/10.1063/1.868284}.
\bibitem[{Cohen et~al.(1991)Cohen, Breuer and Haritonidis}]{Cohen1991}
\bibinfo{author}{Cohen, J.}, \bibinfo{author}{Breuer, K.S.},
  \bibinfo{author}{Haritonidis, J.H.}, \bibinfo{year}{1991}.
\newblock \bibinfo{title}{On the evolution of a wave packet in a laminar
  boundary layer}.
\newblock \bibinfo{journal}{J. Fluid Mech.} \bibinfo{volume}{225},
  \bibinfo{pages}{575--606}.
\newblock \URLprefix \url{http://dx.doi.org/10.1017/S0022112091002185}.
\bibitem[{Craik(1971)}]{Craik1971}
\bibinfo{author}{Craik, A.D.D.}, \bibinfo{year}{1971}.
\newblock \bibinfo{title}{Non-linear resonant instability in boundary layers}.
\newblock \bibinfo{journal}{J. Fluid Mech.} \bibinfo{volume}{50},
  \bibinfo{pages}{393--413}.
\newblock \URLprefix \url{http://dx.doi.org/10.1017/S0022112071002635}.
\bibitem[{Craik(2001)}]{Craik2001}
\bibinfo{author}{Craik, A.D.D.}, \bibinfo{year}{2001}.
\newblock \bibinfo{title}{A model for subharmonic resonance within wavepackets
  in unstable boundary layers}.
\newblock \bibinfo{journal}{J. Fluid Mech.} \bibinfo{volume}{432},
  \bibinfo{pages}{409--418}.
\newblock \URLprefix \url{http://dx.doi.org/10.1017/S0022112001003524}.
\bibitem[{Gaster(1968)}]{Gaster1968}
\bibinfo{author}{Gaster, M.}, \bibinfo{year}{1968}.
\newblock \bibinfo{title}{The development of three-dimensional wave packets in
  a boundary layer}.
\newblock \bibinfo{journal}{J. Fluid Mech.} \bibinfo{volume}{32},
  \bibinfo{pages}{173--184}.
\newblock \URLprefix \url{http://dx.doi.org/10.1017/S0022112068000649}.
\bibitem[{Gaster and Grant(1975)}]{Gaster1975a}
\bibinfo{author}{Gaster, M.}, \bibinfo{author}{Grant, I.},
  \bibinfo{year}{1975}.
\newblock \bibinfo{title}{An experimental investigation of the formation and
  development of a wave packet in a laminar boundary layer}.
\newblock \bibinfo{journal}{Proc. R. Soc. Lond. A.} \bibinfo{volume}{347},
  \bibinfo{pages}{253--269}.
\newblock \URLprefix \url{http://dx.doi.org/10.1098/rspa.1975.0208}.
\bibitem[{Goldstein and Sescu(2008)}]{Goldstein2008}
\bibinfo{author}{Goldstein, M.E.}, \bibinfo{author}{Sescu, A.},
  \bibinfo{year}{2008}.
\newblock \bibinfo{title}{Boundary-layer transition at high free-stream
  disturbance {levels - beyond Klebanoff} modes}.
\newblock \bibinfo{journal}{J. Fluid Mech.} \bibinfo{volume}{613},
  \bibinfo{pages}{95--124}.
\newblock \URLprefix \url{http://dx.doi.org/10.1017/S0022112008003078}.
\bibitem[{Healey(1995)}]{Healey1995}
\bibinfo{author}{Healey, J.J.}, \bibinfo{year}{1995}.
\newblock \bibinfo{title}{A new boundary layer resonance enhanced by wave
  modulation: theory and experiment}.
\newblock \bibinfo{journal}{J. Fluid Mech.} \bibinfo{volume}{304},
  \bibinfo{pages}{231--262}.
\newblock \URLprefix \url{http://dx.doi.org/10.1017/S0022112095004411}.
\bibitem[{Herbert(1984)}]{Herbert1984}
\bibinfo{author}{Herbert, T.}, \bibinfo{year}{1984}.
\newblock \bibinfo{title}{Analysis of the subharmonic route to transition in
  boundary layers}.
\newblock \bibinfo{journal}{AIAA Paper 1984-0009} \URLprefix
  \url{http://dx.doi.org/10.2514/6.1984-9}.
\bibitem[{Herbert(1988)}]{Herbert1988}
\bibinfo{author}{Herbert, T.}, \bibinfo{year}{1988}.
\newblock \bibinfo{title}{Secondary instability of boundary layers}.
\newblock \bibinfo{journal}{Annu. Rev. Fluid Mech.} \bibinfo{volume}{20},
  \bibinfo{pages}{487--526}.
\newblock \URLprefix
  \url{http://www.annualreviews.org/doi/abs/10.1146/annurev.fl.20.010188.002415},
  \DOIprefix\doi{10.1146/annurev.fl.20.010188.002415}.
\bibitem[{Holmes et~al.(2012)Holmes, Lumley, Berkhooz and Rowley}]{Holmes2012}
\bibinfo{author}{Holmes, P.}, \bibinfo{author}{Lumley, J.},
  \bibinfo{author}{Berkhooz, G.}, \bibinfo{author}{Rowley, C.W.},
  \bibinfo{year}{2012}.
\newblock \bibinfo{title}{Turbulence, Coherent Structures, Dynamical Systems
  and Symmetry}.
\newblock \bibinfo{publisher}{Cambridge University Press}.
\bibitem[{Jordinson(1970)}]{Jordinson1970}
\bibinfo{author}{Jordinson, R.}, \bibinfo{year}{1970}.
\newblock \bibinfo{title}{The flat plate boundary layer. part 1. numerical
  integration of the {Orr-Sommerfeld} equation}.
\newblock \bibinfo{journal}{J. Fluid Mech.} \bibinfo{volume}{43},
  \bibinfo{pages}{801--811}.
\newblock \URLprefix \url{http://dx.doi.org/10.1017/S0022112070002756}.
\bibitem[{Kachanov(1987)}]{Kachanov1987}
\bibinfo{author}{Kachanov, Y.S.}, \bibinfo{year}{1987}.
\newblock \bibinfo{title}{On the resonant nature of the breakdown of a laminar
  boundary layer}.
\newblock \bibinfo{journal}{J. Fluid Mech.} \bibinfo{volume}{184},
  \bibinfo{pages}{43--74}.
\newblock \URLprefix \url{http://dx.doi.org/10.1017/S0022112087002805}.
\bibitem[{Kachanov and Levchenko(1984)}]{Kachanov1984}
\bibinfo{author}{Kachanov, Y.S.}, \bibinfo{author}{Levchenko, V.Y.},
  \bibinfo{year}{1984}.
\newblock \bibinfo{title}{The resonant interaction of disturbances at
  laminar-turbulent transition in a boundary layer}.
\newblock \bibinfo{journal}{J. Fluid Mech.} \bibinfo{volume}{138},
  \bibinfo{pages}{209--247}.
\newblock \URLprefix \url{http://dx.doi.org/10.1017/S0022112084000100}.
\bibitem[{Kang(2015)}]{Kang2015a}
\bibinfo{author}{Kang, K.L.}, \bibinfo{year}{2015}.
\newblock \bibinfo{title}{Laminar turbulent transition—spectral analysis and
  coherent structures of DNS wavepackets}.
\newblock Ph.D. thesis. National University of Singapore.
\newblock \URLprefix \url{https://pqdtopen.proquest.com/pubnum/10006031.html}.
\bibitem[{Kang and Yeo(2013)}]{Kang2013}
\bibinfo{author}{Kang, K.L.}, \bibinfo{author}{Yeo, K.S.},
  \bibinfo{year}{2013}.
\newblock \bibinfo{title}{The effect of wavepacket frequency bandwidth on the
  laminar-turbulent transition process in a {Blasius} boundary layer}.
\newblock \bibinfo{journal}{AIAA Paper 2013-2615}
  \DOIprefix\doi{http://dx.doi.org/10.2514/6.2013-2615}.
\bibitem[{Kang and Yeo(2017)}]{Kang2017}
\bibinfo{author}{Kang, K.L.}, \bibinfo{author}{Yeo, K.S.},
  \bibinfo{year}{2017}.
\newblock \bibinfo{title}{Hybrid pod-fft analysis of nonlinear evolving
  coherent structures of dns wavepacket in laminar-turbulent transition}.
\newblock \bibinfo{journal}{Phys. Fluids} \bibinfo{volume}{29},
  \bibinfo{pages}{084105}.
\newblock \URLprefix \url{https://doi.org/10.1063/1.4999348},
  \DOIprefix\doi{10.1063/1.4999348},
  \href{http://arxiv.org/abs/https://doi.org/10.1063/1.4999348}{{\tt
  arXiv:https://doi.org/10.1063/1.4999348}}.
\bibitem[{Kim and Moin(1985)}]{Kim1985}
\bibinfo{author}{Kim, J.}, \bibinfo{author}{Moin, P.}, \bibinfo{year}{1985}.
\newblock \bibinfo{title}{Application of a fractional-step method to
  incompressible {Navier-Stokes} equations}.
\newblock \bibinfo{journal}{J. Comput. Phys.} \bibinfo{volume}{59},
  \bibinfo{pages}{308--323}.
\newblock \URLprefix
  \url{http://www.sciencedirect.com/science/article/pii/0021999185901482},
  \DOIprefix\doi{http://dx.doi.org/10.1016/0021-9991(85)90148-2}.
\bibitem[{Liu and Liu(1994)}]{Liu1994}
\bibinfo{author}{Liu, Z.}, \bibinfo{author}{Liu, C.}, \bibinfo{year}{1994}.
\newblock \bibinfo{title}{Fourth order finite difference and multigrid methods
  for modeling instabilities in flat plate boundary layers\textemdash{2-D} and
  {3-D} approaches}.
\newblock \bibinfo{journal}{Comput. Fluids} \bibinfo{volume}{23},
  \bibinfo{pages}{955--982}.
\newblock \URLprefix
  \url{http://www.sciencedirect.com/science/article/pii/0045793094900639},
  \DOIprefix\doi{http://dx.doi.org/10.1016/0045-7930(94)90063-9}.
\bibitem[{Medeiros(2004)}]{Medeiros2004}
\bibinfo{author}{Medeiros, M.A.F.}, \bibinfo{year}{2004}.
\newblock \bibinfo{title}{The nonlinear evolution of a wavetrain emanating from
  a point source in a boundary layer}.
\newblock \bibinfo{journal}{J. Fluid Mech.} \bibinfo{volume}{508},
  \bibinfo{pages}{287--317}.
\newblock \URLprefix \url{http://dx.doi.org/10.1017/S0022112004009188}.
\bibitem[{Medeiros(2006)}]{Medeiros2006}
\bibinfo{author}{Medeiros, M.A.F.}, \bibinfo{year}{2006}.
\newblock \bibinfo{title}{Nonlinear Wavepackets in Boundary Layers IUTAM
  Symposium on Laminar-Turbulent Transition}. \bibinfo{publisher}{Springer
  Netherlands}. volume~\bibinfo{volume}{78} of \textit{\bibinfo{series}{Fluid
  Mechanics and Its Applications}}.
\newblock pp. \bibinfo{pages}{317--322}.
\bibitem[{Medeiros and Gaster(1999a)}]{Medeiros1999a}
\bibinfo{author}{Medeiros, M.A.F.}, \bibinfo{author}{Gaster, M.},
  \bibinfo{year}{1999}a.
\newblock \bibinfo{title}{The influence of phase on the nonlinear evolution of
  wavepackets in boundary layers}.
\newblock \bibinfo{journal}{J. Fluid Mech.} \bibinfo{volume}{397},
  \bibinfo{pages}{259--283}.
\newblock \URLprefix \url{http://dx.doi.org/10.1017/S0022112099006175}.
\bibitem[{Medeiros and Gaster(1999b)}]{Medeiros1999b}
\bibinfo{author}{Medeiros, M.A.F.}, \bibinfo{author}{Gaster, M.},
  \bibinfo{year}{1999}b.
\newblock \bibinfo{title}{The production of subharmonic waves in the nonlinear
  evolution of wavepackets in boundary layers}.
\newblock \bibinfo{journal}{J. Fluid Mech.} \bibinfo{volume}{399},
  \bibinfo{pages}{301--318}.
\newblock \URLprefix \url{http://dx.doi.org/10.1017/S0022112099006424}.
\bibitem[{Novikov et~al.(2016)Novikov, Egorov and Fedorov}]{Novikov2016}
\bibinfo{author}{Novikov, A.}, \bibinfo{author}{Egorov, I.},
  \bibinfo{author}{Fedorov, A.}, \bibinfo{year}{2016}.
\newblock \bibinfo{title}{Direct numerical simulation of wave packets in
  hypersonic compression-corner flow}.
\newblock \bibinfo{journal}{AIAA Journal} \bibinfo{volume}{54},
  \bibinfo{pages}{2034--2050}.
\newblock \URLprefix \url{https://doi.org/10.2514/1.J054665},
  \DOIprefix\doi{10.2514/1.J054665},
  \href{http://arxiv.org/abs/https://doi.org/10.2514/1.J054665}{{\tt
  arXiv:https://doi.org/10.2514/1.J054665}}.
\bibitem[{Paula et~al.(2010)Paula, W{\"{u}}rz, Kr{\"{a}}mer, Borodulin and
  Kachanov}]{Paula2010}
\bibinfo{author}{Paula, I.B.}, \bibinfo{author}{W{\"{u}}rz, W.},
  \bibinfo{author}{Kr{\"{a}}mer, E.}, \bibinfo{author}{Borodulin, V.I.},
  \bibinfo{author}{Kachanov, Y.S.}, \bibinfo{year}{2010}.
\newblock \bibinfo{title}{Experimental study of resonant interactions of
  modulated waves in a non self-similar boundary layer Seventh IUTAM Symposium
  on Laminar-Turbulent Transition}. \bibinfo{publisher}{Springer Netherlands}.
  volume~\bibinfo{volume}{18} of \textit{\bibinfo{series}{IUTAM Bookseries
  (closed)}}.
\newblock pp. \bibinfo{pages}{549--552}.
\bibitem[{Paula et~al.(2013a)Paula, W{\"{u}}rz, Kr{\"{a}}mer, Borodulin and
  Kachanov}]{Paula2013b}
\bibinfo{author}{Paula, I.B.}, \bibinfo{author}{W{\"{u}}rz, W.},
  \bibinfo{author}{Kr{\"{a}}mer, E.}, \bibinfo{author}{Borodulin, V.I.},
  \bibinfo{author}{Kachanov, Y.S.}, \bibinfo{year}{2013}a.
\newblock \bibinfo{title}{Introduction of Initial Seeds for Subharmonic
  Resonance by Modulation of T-S Waves in an Airfoil Boundary Layer}.
  \bibinfo{publisher}{Springer Berlin Heidelberg}. volume \bibinfo{volume}{121}
  of \textit{\bibinfo{series}{Notes on Numerical Fluid Mechanics and
  Multidisciplinary Design}}. \bibinfo{type}{book
  section}~\bibinfo{chapter}{30}.
\newblock pp. \bibinfo{pages}{245--252}.
\bibitem[{Paula et~al.(2013b)Paula, W{\"{u}}rz, Kr{\"{a}}mer, Borodulin and
  Kachanov}]{Paula2013}
\bibinfo{author}{Paula, I.B.}, \bibinfo{author}{W{\"{u}}rz, W.},
  \bibinfo{author}{Kr{\"{a}}mer, E.}, \bibinfo{author}{Borodulin, V.I.},
  \bibinfo{author}{Kachanov, Y.S.}, \bibinfo{year}{2013}b.
\newblock \bibinfo{title}{Weakly nonlinear stages of boundary-layer transition
  initiated by modulated tollmien schlichting waves}.
\newblock \bibinfo{journal}{J. Fluid Mech.} \bibinfo{volume}{732},
  \bibinfo{pages}{571--615}.
\newblock \URLprefix \url{http://dx.doi.org/10.1017/jfm.2013.420}.
\bibitem[{Rhie and Chow(1983)}]{Rhie1983}
\bibinfo{author}{Rhie, C.M.}, \bibinfo{author}{Chow, W.L.},
  \bibinfo{year}{1983}.
\newblock \bibinfo{title}{Numerical study of the turbulent flow past an airfoil
  with trailing edge separation}.
\newblock \bibinfo{journal}{AIAA J.} \bibinfo{volume}{21},
  \bibinfo{pages}{1525--1532}.
\newblock \URLprefix \url{http://dx.doi.org/10.2514/3.8284}.
\bibitem[{Sengupta et~al.(2012)Sengupta, Bhaumik and Bhumkar}]{Sengupta2012}
\bibinfo{author}{Sengupta, T.K.}, \bibinfo{author}{Bhaumik, S.},
  \bibinfo{author}{Bhumkar, Y.G.}, \bibinfo{year}{2012}.
\newblock \bibinfo{title}{Direct numerical simulation of two-dimensional
  wall-bounded turbulent flows from receptivity stage}.
\newblock \bibinfo{journal}{Phys. Rev. E} \bibinfo{volume}{85},
  \bibinfo{pages}{026308}.
\newblock \URLprefix \url{https://link.aps.org/doi/10.1103/PhysRevE.85.026308},
  \DOIprefix\doi{http://dx.doi.org/10.1103/PhysRevE.85.026308}.
\bibitem[{Sengupta et~al.(2006a)Sengupta, Rao and
  Venkatasubbaiah}]{Sengupta2006a}
\bibinfo{author}{Sengupta, T.K.}, \bibinfo{author}{Rao, A.K.},
  \bibinfo{author}{Venkatasubbaiah, K.}, \bibinfo{year}{2006}a.
\newblock \bibinfo{title}{Spatio-temporal growth of disturbances in a boundary
  layer and energy based receptivity analysis}.
\newblock \bibinfo{journal}{Phys. Fluids} \bibinfo{volume}{18},
  \bibinfo{pages}{094101}.
\newblock \URLprefix
  \url{http://scitation.aip.org/content/aip/journal/pof2/18/9/10.1063/1.2348732},
  \DOIprefix\doi{http://dx.doi.org/10.1063/1.2348732}.
\bibitem[{Sengupta et~al.(2006b)Sengupta, Rao and
  Venkatasubbaiah}]{Sengupta2006}
\bibinfo{author}{Sengupta, T.K.}, \bibinfo{author}{Rao, A.K.},
  \bibinfo{author}{Venkatasubbaiah, K.}, \bibinfo{year}{2006}b.
\newblock \bibinfo{title}{Spatiotemporal growing wave fronts in spatially
  stable boundary layers}.
\newblock \bibinfo{journal}{Phys. Rev. Lett.} \bibinfo{volume}{96},
  \bibinfo{pages}{224504}.
\newblock \URLprefix \url{http://dx.doi.org/10.1103/PhysRevLett.96.224504}.
\bibitem[{Sharma et~al.(2018)Sharma, Sengupta and Bhaumik}]{Sharma2018}
\bibinfo{author}{Sharma, P.}, \bibinfo{author}{Sengupta, T.K.},
  \bibinfo{author}{Bhaumik, S.}, \bibinfo{year}{2018}.
\newblock \bibinfo{title}{Three-dimensional transition of
  zero-pressure-gradient boundary layer by impulsively and nonimpulsively
  started harmonic wall excitation}.
\newblock \bibinfo{journal}{Phys. Rev. E} \bibinfo{volume}{98},
  \bibinfo{pages}{053106}.
\newblock \URLprefix \url{https://link.aps.org/doi/10.1103/PhysRevE.98.053106},
  \DOIprefix\doi{10.1103/PhysRevE.98.053106}.
\bibitem[{Smith and Stewart(1987)}]{Smith1987}
\bibinfo{author}{Smith, F.T.}, \bibinfo{author}{Stewart, P.A.},
  \bibinfo{year}{1987}.
\newblock \bibinfo{title}{The resonant-triad nonlinear interaction in
  boundary-layer transition}.
\newblock \bibinfo{journal}{J. Fluid Mech.} \bibinfo{volume}{179},
  \bibinfo{pages}{227–252}.
\newblock \DOIprefix\doi{10.1017/S0022112087001502}.
\bibitem[{Sundaram et~al.(2019)Sundaram, Sengupta and Sengupta}]{Sundaram2019}
\bibinfo{author}{Sundaram, P.}, \bibinfo{author}{Sengupta, T.K.},
  \bibinfo{author}{Sengupta, S.}, \bibinfo{year}{2019}.
\newblock \bibinfo{title}{Is tollmien-schlichting wave necessary for transition
  of zero pressure gradient boundary layer flow?}
\newblock \bibinfo{journal}{Physics of Fluids} \bibinfo{volume}{31},
  \bibinfo{pages}{031701}.
\newblock \URLprefix \url{https://doi.org/10.1063/1.5089294},
  \DOIprefix\doi{10.1063/1.5089294},
  \href{http://arxiv.org/abs/https://doi.org/10.1063/1.5089294}{{\tt
  arXiv:https://doi.org/10.1063/1.5089294}}.
\bibitem[{Temam(1984)}]{Temam1984}
\bibinfo{author}{Temam, R.}, \bibinfo{year}{1984}.
\newblock \bibinfo{title}{{Navier-Stokes} Equations: Theory and Numerical
  Analysis}. volume \bibinfo{volume}{343}.
\newblock \bibinfo{publisher}{AMS Chelsea Publishing}.
\bibitem[{Wang(2003)}]{Wang2003}
\bibinfo{author}{Wang, Z.Y.}, \bibinfo{year}{2003}.
\newblock \bibinfo{title}{Computational simulation of unsteady boundary layer
  over compliant surfaces}.
\newblock Ph.D. thesis. National University of Singapore.
\newblock \URLprefix \url{http://scholarbank.nus.sg/handle/10635/13874}.
\bibitem[{Wang et~al.(2005)Wang, Yeo and Khoo}]{Wang2005}
\bibinfo{author}{Wang, Z.Y.}, \bibinfo{author}{Yeo, K.S.},
  \bibinfo{author}{Khoo, B.C.}, \bibinfo{year}{2005}.
\newblock \bibinfo{title}{Spatial direct numerical simulation of transitional
  boundary layer over compliant surfaces}.
\newblock \bibinfo{journal}{Comput. Fluids} \bibinfo{volume}{34},
  \bibinfo{pages}{1062--1095}.
\newblock \URLprefix
  \url{http://www.sciencedirect.com/science/article/pii/S0045793004001318},
  \DOIprefix\doi{http://dx.doi.org/10.1016/j.compfluid.2004.08.005}.
\bibitem[{Wesseling and Oosterlee(2001)}]{Wesseling2001}
\bibinfo{author}{Wesseling, P.}, \bibinfo{author}{Oosterlee, C.W.},
  \bibinfo{year}{2001}.
\newblock \bibinfo{title}{Geometric multigrid with applications to
  computational fluid dynamics}.
\newblock \bibinfo{journal}{J. Comput. Appl. Math.} \bibinfo{volume}{128},
  \bibinfo{pages}{311--334}.
\newblock \URLprefix
  \url{http://www.sciencedirect.com/science/article/pii/S0377042700005173},
  \DOIprefix\doi{http://dx.doi.org/10.1016/S0377-0427(00)00517-3}.
\bibitem[{Wu and Choudhari(2001)}]{Wu2001}
\bibinfo{author}{Wu, X.}, \bibinfo{author}{Choudhari, M.},
  \bibinfo{year}{2001}.
\newblock \bibinfo{title}{Effects of long-wavelength {Klebanoff} modes on
  boundary-layer instability}, in: \bibinfo{booktitle}{Annual Research Briefs
  2001}. \bibinfo{publisher}{Center for Turbulence Research, Stanford
  University}.
\bibitem[{W{\"{u}}rz et~al.(2012)W{\"{u}}rz, Sartorius, Kloker, Borodulin,
  Kachanov and Smorodsky}]{Wurz2012a}
\bibinfo{author}{W{\"{u}}rz, W.}, \bibinfo{author}{Sartorius, D.},
  \bibinfo{author}{Kloker, M.}, \bibinfo{author}{Borodulin, V.I.},
  \bibinfo{author}{Kachanov, Y.S.}, \bibinfo{author}{Smorodsky, B.V.},
  \bibinfo{year}{2012}.
\newblock \bibinfo{title}{Detuned resonances of {Tollmien--Schlichting} waves
  in an airfoil boundary layer: Experiment, theory, and direct numerical
  simulation}.
\newblock \bibinfo{journal}{Phys. Fluids} \bibinfo{volume}{24},
  \bibinfo{pages}{094103--25}.
\newblock \URLprefix
  \url{http://scitation.aip.org/content/aip/journal/pof2/24/9/10.1063/1.4751246},
  \DOIprefix\doi{http://dx.doi.org/10.1063/1.4751246}.
\bibitem[{Yeo et~al.(2010)Yeo, Zhao, Wang and Ng}]{Yeo2010}
\bibinfo{author}{Yeo, K.S.}, \bibinfo{author}{Zhao, X.}, \bibinfo{author}{Wang,
  Z.Y.}, \bibinfo{author}{Ng, K.C.}, \bibinfo{year}{2010}.
\newblock \bibinfo{title}{{DNS} of wavepacket evolution in a {Blasius} boundary
  layer}.
\newblock \bibinfo{journal}{J. Fluid Mech.} \bibinfo{volume}{652},
  \bibinfo{pages}{333--372}.
\newblock \URLprefix \url{http://dx.doi.org/10.1017/S0022112009994095}.
\bibitem[{Zhao(2007)}]{Zhao2007}
\bibinfo{author}{Zhao, X.}, \bibinfo{year}{2007}.
\newblock \bibinfo{title}{Computational simulation of wavepacket evolution over
  compliant surfaces}.
\newblock Ph.D. thesis. National University of Singapore.
\newblock \URLprefix \url{http://scholarbank.nus.edu.sg/handle/10635/16155}.

\end{thebibliography}

\end{document}